\documentclass[reprint,amsmath,twocolumn,amssymb, nobibnotes, aps, pre, superscriptaddress]{revtex4-1}

\usepackage{graphicx}
\usepackage{dcolumn}
\usepackage{bm}
\usepackage{color}
\usepackage{tikz}
\usetikzlibrary{shapes}
\usetikzlibrary{backgrounds}
\usetikzlibrary{plotmarks}

\begin{document}
\def \beq{\begin{equation}}
\def \eeq{\end{equation}}
\def \bea{\begin{eqnarray}}
\def \eea{\end{eqnarray}}
\def \bem{\begin{displaymath}}
\def \eem{\end{displaymath}}
\def \P{\Psi}
\def \Pd{|\Psi(\boldsymbol{r})|}
\def \Pds{|\Psi^{\ast}(\boldsymbol{r})|}
\def \Po{\overline{\Psi}}
\def \bs{\boldsymbol}
\def \bl{\bar{\boldsymbol{l}}}
\newcommand{\ihat}{\hat{\textbf{\i}}}
\newcommand{\jhat}{\hat{\textbf{\j}}}

\title{Peierls-Nabarro barrier  {effect in} nonlinear  {Floquet} topological insulators}

\author{Mark J. Ablowitz}
\affiliation{Department of Applied Mathematics, University of Colorado, Boulder, Colorado, USA}
\author{Justin T. Cole}
\affiliation{Department of Mathematics, University of Colorado, Colorado Springs, Colorado, USA}
\author{Pipi Hu}
\affiliation{Yau Mathematical Sciences Center, Tsinghua University, Beijing, China}
\affiliation{Yanqi Lake Beijing Institute of Mathematical Sciences and Applications, Beijing, China }
\author{Peter Rosenthal}
\affiliation{Department of Applied Mathematics, University of Colorado, Boulder, Colorado, USA}
\date{\today}     

\begin{abstract}
The Peierls-Nabarro barrier is a discrete effect {that frequently occurs in discrete} nonlinear 
systems. A signature of the barrier is the slowing and eventual stopping of discrete solitary waves. This work examines intense electromagnetic waves propagating through a periodic  {honeycomb} lattice of helically-driven waveguides{, which serves as a paradigmatic  Floquet topological insulator.}  
{Here} it is shown that 
 {discrete} topologically protected edge modes {\it do not} suffer from the typical slowdown associated with the Peierls-Nabarro barrier. Instead{, as a result of their topological nature}, the modes always move forward {and} 
 redistribute their energy: a narrow (discrete) mode transforms into a wide effectively continuous mode. 
{On the other hand}, a discrete edge mode that is {\it not} topologically protected {\it does} eventually  {slow down and 
{stop}} propagating. 
{Topological modes that are initially narrow modes naturally tend to wide envelope states {that} 
are described} 
by a generalized nonlinear Schr\"odinger equation. These results {provide insight into} the nature of nonlinear topological insulators and their application.

\end{abstract}
\maketitle

\section{Introduction}
\label{intro}


The study of  topological insulators and topologically protected modes has captured wide attention.
A typical topological insulator is conductive along the edge or surface of the system and insulating in the interior or bulk of the material. The edge modes are exponentially localized near the boundary and are remarkably robust against local defects and disorder. The edge modes supported by these systems are known as topologically protected states and they are associated with nontrivial topological invariants of the bulk eigenmodes. Topological invariants are a common way of identifying and characterizing  topologically protected modes.

{In the case of Chern insulators,  topologically protected modes are associated with a nontrivial Chern number of the bulk states. Moreover, 
these  systems support nonreciprocal (unidirectional) edge modes that do not backscatter {from} lattice defects along a domain boundary. 
{One can realize a Chern insulator by} {breaking time-reversal symmetry.} 
This paper focuses on a Chern insulator system.}


{T}he first experimental realization of a topological insulator in an electromagnetic system was a magneto-optic medium with an external magnetic field applied to an array of ferrite rods {\cite{Wang08,Wang09,Lu14}.} Magnet-free topological insulators have been constructed in other systems {such as} 
coupled ring resonators \cite{Hafzei13} and helically-varying waveguide arrays \cite{Rechtsman13}.  {Many} other photonic topological insulators have {also} been proposed theoretically and experimentally {--e.g. see the reviews \cite{Ozawa19,Khanikaev17}.}


{Most studies of} topological insulators {have involved} linear systems. In the linear regime, introducing periodic media that breaks time-reversal symmetry 
{has led to} nonreciprocal modes with nontrivial Chern numbers  \cite{haldane08}.
{{Meanwhile{,}} 
 investigations of {\it nonlinear} topological insulators {(TIs)} {are} 
less developed  {in comparison cf.} 
\cite{Rechtsman13, Ablowitz20}.} {A {natural} question {to ask is whether} 
the principles of topological protection extend to the nonlinear regime.
{A} system that has been {carefully} studied is the propagation of intense (nonlinear) electromagnetic waves in Kerr media with an array of 
waveguides that are  helically driven in the direction of beam propagation. The dynamics of these waves are described by the paraxial wave equation, which is a nonlinear Schr\"odinger (NLS) equation with periodic potential {\cite{Lumer13,Ablowitz14, Ablowitz17, Ablowitz19}.}

{On a boundary edge}, {intense electric fields are} 
described by a (linear) topologically protected mode modulated by a slowly varying envelope that satisfies a generalized NLS equation \cite{Ablowitz14, Ablowitz17}. As a result, these systems exhibit many classic NLS features, such as modulational instability \cite{Lumer16} and soliton solutions \cite{Ablowitz14, Ablowitz17}. {These edge waves inherit the  topological properties of the corresponding  linear modes;
{{i.e.} 
they are unidirectional and do not reflect off defects.} 
{Additionally, i}n the bulk, there are nonlinear {topological} modes that propagate with a cyclotronic motion \cite{Lumer13,Mukherjee20}.}

{Semi-discrete wave equations occur widely in physics and mathematics; e.g. discrete soliton propagation in optical waveguide arrays \cite{ChristJos88} and numerical approximations of continuous partial differential equations by their discrete counterparts. {Traveling waves are ubiquitous in continuous systems.} 
But under discretization, the existence of uniformly  traveling waves is not at all clear.  Discrete waves {can} suffer from what is commonly referred to as {the} Peierls-Nabarro (PN) energy barrier{, which is the amount of energy needed to travel one lattice  site. The source of the energy barrier is a position dependent value of the Hamiltonian (see below)} 
{e.g.} 
whether the `center' of the wave is at a 
{grid} point or not. The PN energy barrier prevents the discrete wave from having the translation invariance commonly found in continuous-wave systems.
{As a result, the PN barrier  can slow down and eventually stop 
discrete localized traveling waves. }

On the contrary{,} our studies reveal that unlike their nontopological counterparts{,} topologically protected {edge wave} envelopes do not slow down or stop.
In the topological case, an initially discrete mode rearranges its spatial profile into a wide, effectively continuous, form. Furthermore, over long time scales these nonlinear modes are found to approach {wide nonlinear envelopes, which} 
are governed by a generalized NLS equation. 
Although we have directed our attention to a particular honeycomb lattice system, this novel effect of the PN barrier  should be observable in many topological systems as well as in the laboratory. 
{This is a purely nonlinear phenomenon; linear modes have no PN barrier effect, but they do suffer from dispersion-induced decay.}

{We consider moderate to extremely narrow initial {edge} wave profiles. Both {edge} wave profiles shed radiation and tend to modes governed by NLS-type equations. If we begin with moderately narrow widths, then after a long time evolution  we find modes that are approximated by solitons of the classical NLS equation. On the other hand, in  the  extremely narrow case, 
the long-time modes are approximated by the third-order dispersive NLS equation, whose modes continue to shed radiation during evolution. This is described in terms of an underlying Hamiltonian of the {modes.}} 

\subsection{Background}
\label{background_subsection}

{There is a large literature that discusses traveling waves in discrete systems and whether or not  discrete traveling waves  exist. We will not present a review here. Rather we only discuss some well-known results as a means of comparison with {our findings in a topological system below.} }

An important early study of the PN barrier  
considered kink 
modes of the {semi-}discrete sine-Gordon {(DSG)} equation \cite{Peyrard84}  {
\begin{equation}
\label{dSG}
\frac{d^2u_n}{dt^2}=  \frac{u_{n+1}-2u_n + u_{n-1}}{h^2} - \sin u_n ~ , 
\end{equation}
where $u_n  = u(nh)$ for $h >0$ and $n \in \mathbb{Z}$}  with the corresponding Hamiltonian
\begin{align}
& H_{\rm DSG} =\\
\nonumber & \sum_n \frac{1}{2}  \left( \frac{d u_n}{dt} \right)^2 + \frac{1}{2 h^2} \left( u_{n+1} - u_n\right)^2 +  \left( 1 - \cos u_n \right) ~,
\end{align}
{ and whose continuum limit {($h \rightarrow 0$)}} {yields} 
\begin{equation}
\label{SG}
\frac{\partial^2 u}{\partial t^2} =  \frac{\partial^2 u}{\partial x^2} - \sin u ~ .
\end{equation}
 There it was found that modes whose kink width was on the order of the grid spacing were affected by the PN barrier. {Consider the traveling soliton solution of (\ref{SG})
\begin{equation}
\label{SG_soliton}
u_D(x,t) = 4 \arctan\left( \exp\left[ \frac{x - vt - x_0}{\sqrt{1 - v^2}}  \right] \right) ~,
\end{equation}
which is initially centered at $ x_0$ and moves with constant velocity $v$. Now examine this kink solution in DSG equation (\ref{dSG}) with initial condition $u_n(0) = u_D(x_n,0)$ such that $x_n = n h$ with grid spacing $h = 1$.} Kinks centered at lattice sites (on-site) {have} {higher} 
energy (Hamiltonian) than kink solutions with two lattice sites equidistant  from the center {(off-site)}-- see Fig.~\ref{DSG_IC}.

Discrete kinks with different energies cannot propagate uniformly.
It was found that after propagation for some time{, these discrete} traveling waves  slowed down and eventually approached a steady-state {with a distinct and identifiable velocity}. The transition was accompanied by a significant amount of radiation shedding from the mode.

\begin{figure}
\centering
\includegraphics[scale=.4]{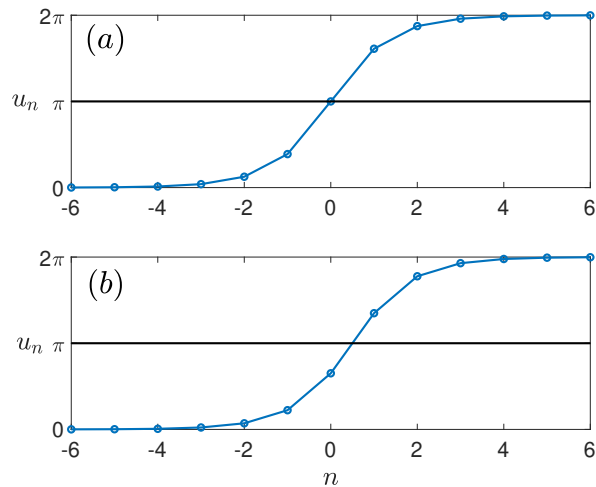}
 \caption{{ {Continuous kink  solution (\ref{SG_soliton}) evaluated on a discrete grid with $v = 1/2$ and $h = 1$.} 
 (a) ``On-site" mode centered at grid point $n = 0$ {where $x_0 = 0$}; (b) ``Off-site" mode {($x_0 = 1/2$)}, {\it not} centered at any grid point.} \label{DSG_IC}}
\end{figure}

A few years later two different discrete NLS systems were studied  numerically and perturbatively \cite{Kivshar93}{:} the `standard' discrete NLS (DNLS) equation {
\begin{equation}
\label{stdNLS}
i\frac{du_n}{dt}+ \frac{u_{n+1}+ u_{n-1}-2u_n}{h^2}+2|u_n|^2u_n=0 ~,
\end{equation}
} and an integrable discretization of NLS (IDNLS) 
\begin{equation}
\label{INLS}
i\frac{du_n}{dt}+ \frac{u_{n+1}+ u_{n-1}-2u_n}{h^2}+|u_n|^2(u_{n+1}+ u_{n-1})=0 ~.
\end{equation}
{The Hamiltonian corresponding to DNLS (\ref{stdNLS}) is given by
\begin{align}
&H_{\rm DNLS} = \\ \nonumber
& -\sum_n \left(  \frac{1}{h^2}  \left[ u_n \left( u_{n+1}^* - u_n^* \right) + u_n^* \left(  u_{n+1} - u_n  \right) \right] 
  +  |u_n|^4  \right)~ ,
\end{align}
and the Hamiltonian of IDNLS (\ref{INLS}) is {
\begin{align}
&H_{\rm IDNLS} = \\ \nonumber
&-  \sum_n \left( \frac{1}{h^2}   u_n^* \left( u_{n+1} + u_{n-1} \right)  
  - \frac{2}{h^4} \log\left(1 + h^2 |u_n|^2 \right)  \right)~ ,
\end{align}
} where $^*$ denotes complex conjugation.

{In the context of the PN barrier, the main difference between these two systems is that the IDNLS Hamiltonian is translation-invariant, while DNLS is not. Certain localized waves of (\ref{stdNLS}) do suffer from the PN barrier (see below).} {The reader can find reviews and a discussion of localized waves traveling at finite wave speeds in
discrete NLS type systems in \cite{Eilbeck1993,Eilbeck2003,Kevrikides2009,Rothos2016,Rothos2005} and references therein. Both Hamiltonians are {time-independent.} } In the continuum limit both of these systems reduce to the cubic NLS equation
\begin{equation}
\label{NLS}
i\frac{\partial u}{\partial t}+ \frac{\partial^2 u}{\partial x^2}+2|u|^2u=0 ~.
\end{equation}
{This equation} admits traveling soliton solutions 
\begin{equation}
\label{soliton_soln}
u_S(x,t) = \eta ~\rm{sech}\left[ \eta (x + 2 \xi t - x_0) \right] e^{- i \xi x - i(\xi^2 - \eta^2) t} ~,
\end{equation}
with initial position $x_0$, peak magnitude $\eta$, and  velocity $ -2\xi$. }

{The DNLS equation can exhibit behavior that resembles 
DSG equation; namely 
solitary waves centered at a grid point vs. {those} centered between sites have different energies.
 In \cite{Kivshar93} this energy difference was referred to as the PN barrier, motivated by a similar energy barrier in solid state physics \cite{Peierls1940,Nabarro1947}. Consider discretizing in space by $x_n = n h$ with $h = 1$.  In Fig.~\ref{DNLS_IC} solution (\ref{soliton_soln}) centered at $x_0$ is evaluated on a discrete grid by $u_n = u_S(n,0)$.  {Unlike} the DSG modes, the centered ``on-site'' profile has a lower energy compared to the ``off-site'' arrangement; that is $H_{\rm DNLS}^{\rm on} < H_{\rm DNLS}^{\rm off}.$ In \cite{Flach1999} it was shown that traveling waves of the elementary form given by $u_S(nh,t)$ do {\it not} exist for DNLS (\ref{stdNLS}). On the other hand, the integrable Hamiltonian is invariant and $H_{\rm IDNLS}^{\rm on} = H_{\rm IDNLS}^{\rm off}.$} 

\begin{figure}
\centering
\includegraphics[scale=.4]{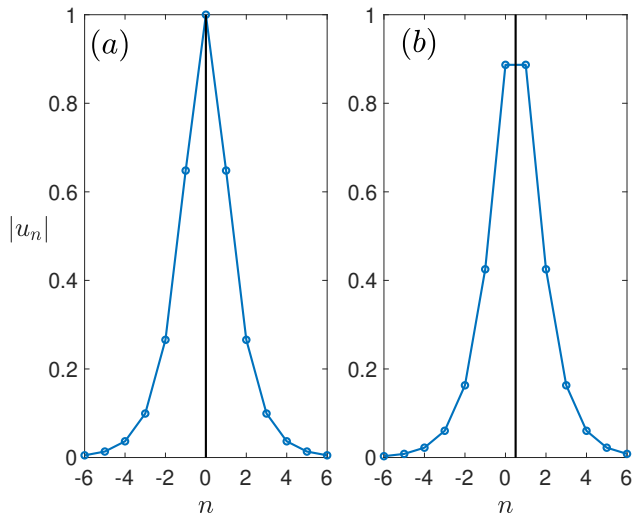}
 \caption{{Initial solitons (\ref{soliton_soln}) evaluated on a discrete grid with $\xi = -1, \eta = 1,$ and $h = 1$.} 
 (a) ``On-site" mode {($x_0 = 0$)} centered at grid point $n = 0$; (b) ``Off-site" mode {($x_0 = 1/2$)}  {\it not} centered at any grid point. \label{DNLS_IC}}
\end{figure}

To illuminate the PN barrier effect, {let us} consider the {dynamics of DNLS. 
Taking spatial grid $x_n = n h$, where} 
 $h = 1, \xi = -1,$ and {$x_0 = -250,$} we evolve (\ref{stdNLS}) with initial data $u_n(0) = u_S(x_n,0)$ for different values of $\eta$. When $0 < \eta \ll 1$ the profile is short and wide (effectively continuous), however when $\eta \gg 1$ the profile is highly discrete (and nonlinear). 
The peak location of the soliton magnitude ($\max_n |u_n|$) is tracked and shown in Fig.~\ref{DNLS_peak_pos}. As $\eta $ increases, the PN energy barrier increases causing {the} mode to slow down and eventually stop propagating altogether. We note that for small $\eta$ it takes a very long time (exponentially long time) for the wave to slow down.

\begin{figure}
\centering
\includegraphics[scale=.43]{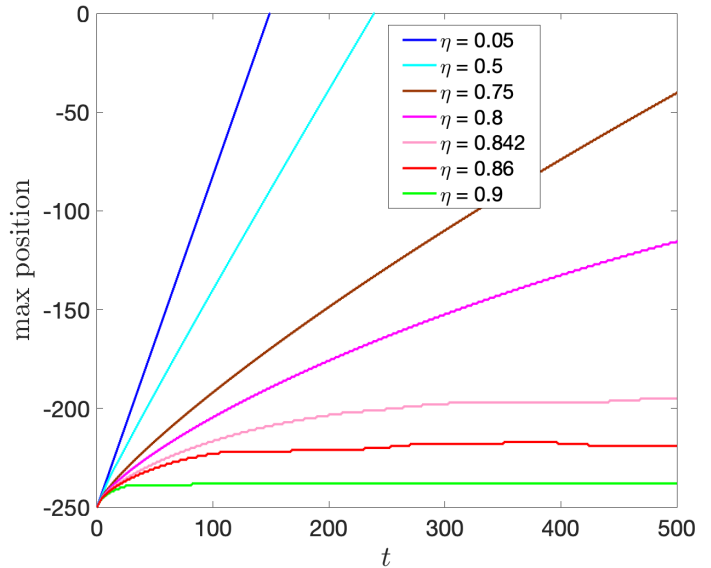}
 \caption{{Location of discrete soliton peak as a function of time for the DNLS Eq.~(\ref{stdNLS}) with $h = 1$. Initial conditions are taken from (\ref{soliton_soln}) with $\xi = -1,$  {$x_0 = -250$} and different values of $\eta$. At fixed $t$,  curves appear top-to-bottom in increasing values of $\eta$}. \label{DNLS_peak_pos}}
\end{figure}

On the other hand{,} for the integrable case {(\ref{INLS})} {its soliton}  {solution} does not have differing energies whether the {center is on} 
or off discrete lattice sites. Indeed the solution {shows that the discretization parameter may be considered as a continuous variable} \cite{Ablowitz_Discrete_Book}.  {In \cite{Kivshar93} the DNLS equation was considered as a perturbation of the  IDNLS equation.}
 In \cite{Morandotti99} the effects of the PN barrier were experimentally observed in a nonlinear optical waveguide system.

 Subsequently, a different perturbation method \cite{MJAZM2003} using  soliton perturbation theory on the continuous NLS equation \cite{Karpman98} showed similar results for the DNLS equation: the change in amplitude and speed from the continuous mode is exponentially small in the discretization parameter; see also \cite{Jenkinson16,Jenkinson17} for a careful discussion of the PN barrier in local and nonlocal DNLS equations in multiple dimensions. {Finally, we point out that} the existence of supersonic  traveling waves associated with certain discrete Fermi-Pasta-Ulam and nonlocal Klein-Gordon equations have also been considered; cf. \cite{pego1999,bates2006}.\\

\subsection{Discrete model of nonlinear topological insulator}
\label{HC_TI}
{We} {now consider} 
a discrete model describing the behavior of an electric field envelope propagating through a honeycomb waveguide array that is helically-varying in the longitudinal direction. This model is inspired by the experiments reported in \cite{Rechtsman13}. 

In nondimensional units, the unit cell of a honeycomb lattice is a parallelogram defined by the lattice vectors 
\begin{equation}
\label{lattice_vecs}
{\bf v}_1 =  \frac{\sqrt{3}}{2} \left( \sqrt{3} , 1\right) \; , \; {\bf v}_2 =  \frac{\sqrt{3}}{2} \left( \sqrt{3} , -1 \right) \; ,
\end{equation}
each of length $\sqrt{3}$.
Each unit cell contains two lattice sites ($a$ {sites} and $b$ {sites}) separated by the vector ${\bf d} = (1,0)${; see Fig. \ref{honey_lattice_fig}.} Below we express the lattice vectors as ${\bf v}_1 = {\bf w}_1 + {\bf w}_2$ and ${\bf v}_2 = {\bf w}_2 - {\bf w}_1$, where ${\bf w}_1  = \frac{\sqrt{3}}{2} (0,1)$ and ${\bf w}_2  = \frac{3}{2} (1,0)$. As a result, the $a$ lattices sites are located at points $\left\{ {\bf v}_a | {\bf v}_a = m {\bf w}_1 + n{\bf w}_2 - {\bf d} \right\}$ and the $b$ sites at $\left\{ {\bf v}_b | {\bf v}_b = m {\bf w}_1 + n{\bf w}_2  \right\}$ where $m,n \in \mathbb{Z}$. A diagram of the honeycomb lattice indexed in terms of the vectors ${\bf w}_1$ and ${\bf w}_2$ is shown in Fig.~\ref{honey_lattice_fig}. This change of variable makes it easier to discuss edge modes that propagate in the $x$ and $y$ directions. Note that the index $m (n)$ corresponds to shifts in the $y (x)$ direction.

\begin{figure}
\centering
\includegraphics[scale=.8]{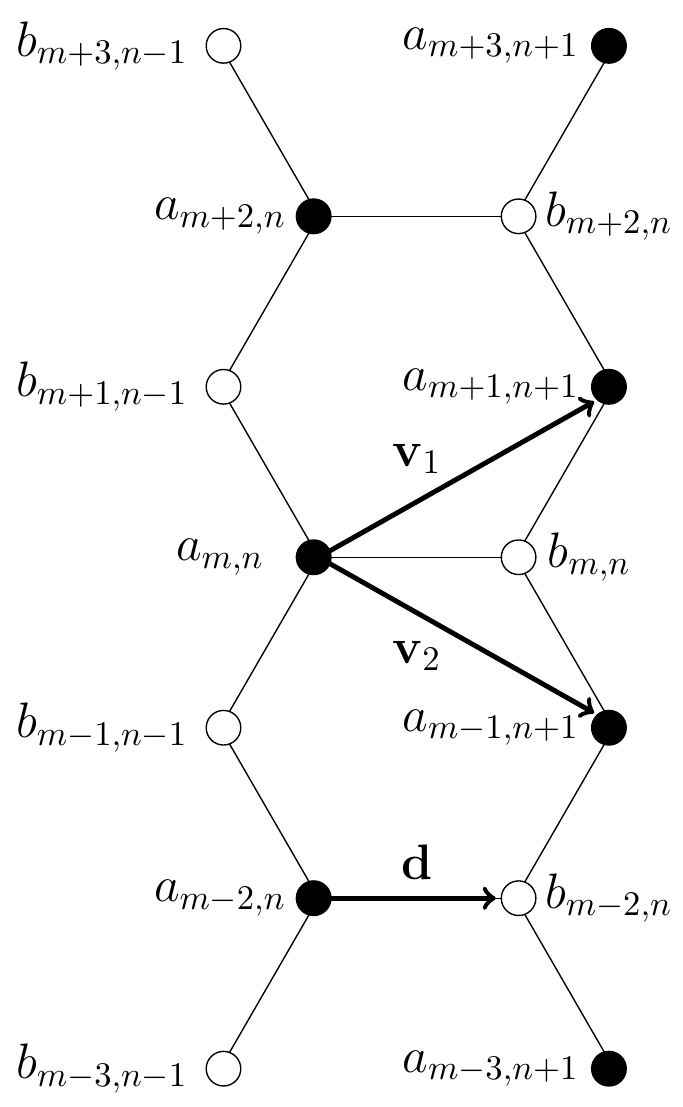}
 \caption{Honeycomb lattice with {lattice} vectors ${\bf v}_1, {\bf v}_2$ defined in (\ref{lattice_vecs}). 
 {The} zig-zag (armchair) edge is parallel to the $y~(x)$ axis. \label{honey_lattice_fig}}
\end{figure}

A helically-varying honeycomb lattice is described by the semi-discrete {nonlinear} system of equations {(see \cite{Ablowitz14,Ablowitz17})}
\begin{align}
\label{HC_TB_eqn1}
& i \frac{d a_{mn}}{dz} + [\mathcal{L}_+ b]_{mn} + \sigma |a_{mn}|^2 a_{mn} = 0 \; , \\
\label{HC_TB_eqn2}
& i \frac{d b_{mn}}{dz} + [\mathcal{L}_- a]_{mn} + \sigma |b_{mn}|^2 b_{mn} = 0 \; ,
\end{align}
where $\sigma \ge 0$ and 
\begin{align*}
 [\mathcal{L}_+ b]_{mn}   =&  e^{i \theta_0(z)} b_{mn}  \\ \nonumber
+ &  \rho \left[ e^{i \theta_1(z)}b_{m-1,n-1} + e^{i \theta_2(z)} b_{m+1,n-1} \right] \; , \\
 [\mathcal{L}_- a]_{mn}   =&  e^{-i \theta_0(z)} a_{mn}  \\ \nonumber
+ &  \rho \left[ e^{-i \theta_1(z)}a_{m+1,n+1} + e^{-i \theta_2(z)} a_{m-1,n+1} \right] \; ,
\end{align*}
such that $\rho > 0,$ $\theta_j(z) = ({\bf d} - {\bf v}_j) \cdot {\bf A}(z)$ for $j = 0,1,2$ and ${\bf v}_ 0 = {\bf 0}.$ The vector potential {induced by the helical variation} 
is ${\bf A}(z) = \kappa \left( \sin(z/\epsilon) , - \cos(z/\epsilon) \right)$, where $\kappa \ge 0 $ is proportional to the helix radius. 
The inverse angular frequency is given by $\epsilon \approx 0.245$  
with a helix pitch of $T = 2 \pi \epsilon = 1.54.$ We point out that the linear coefficients are periodic in $z$: $\theta_j(z + T) = \theta_j(z)$.

{With} regard to {changing} the topology of the system, the crucial parameter is $\rho$ which corresponds to asymmetry in the waveguide profiles: $\rho = 1$ are circular waveguides, $\rho < (>) 1$  are elliptical waveguides with major axis parallel to the $x (y)$ direction. 
{By adjusting} this parameter we can induce a topological transition in the system. Specifically, moving from $\rho: 1 \rightarrow 0$, it is possible to undergo a topological transition from a nontrivial to a trivial {topological} state.
 
To ultimately understand the nonlinear problem, we first study two {linear} 
{reductions} in detail: (a) the bulk (infinite) case, and (b) the edge (semi-infinite) case. In the former {case} we take a two-dimensional Fourier modes and compute the bulk-dispersion surfaces. In the latter case, we take the finite dimension to be the $x$-direction with zig-zag boundaries {parallel to $y$}. At the zig-zag edge we impose Dirichlet zero boundary conditions, and in the infinite $y$-direction we take a 1D Fourier mode. Physically, the zero boundary conditions model an overall absence of the beam outside the waveguide array. 

\subsection{Paper outline}
\label{outline_paper}
 
 Having introduced our model, we now outline the paper. In Sec.~\ref{linear_bands}, the bulk and edge linear Floquet-dispersion bands are computed. Depending on the structure of the waveguides, the eigenmodes associated with these bands may possess nontrivial Chern numbers. In Sec.~\ref{asymptotics_sec}, an asymptotic analysis is performed in a rapidly-oscillating regime. From it comes a description of weakly nonlinear modes known as {\it topologically protected edge solitons}. In Sec.~\ref{PN_section}, numerical simulations reveal that discrete, topologically protected modes do not stop. 
On the other hand, nontopological modes eventually slow down and stop much like they {can} 
for {DNLS.} 
{In Sec.~\ref{long_time_sec}, 
we examine the character of the topological modes after 
long distances.}  We conclude in Sec.~\ref{conclude_sec}.

\section{Floquet Bands and Chern Invariants}
\label{linear_bands}

Before proceeding to the nonlinear problem, it is beneficial to first understand the linear ($\sigma = 0$) case. Physically, this corresponds to low-intensity fields where $|a_{mn}|^2, |b_{mn}|^2 \ll 1.$ In this section, we look for Floquet modes and derive insight from their corresponding band diagrams. 

First, the bulk problem is studied. The spectral surfaces are computed for a typical set of values. Two topological regimes are identified: (a) trivial for $\rho < \rho^*$ and (b) nontrivial for $\rho > \rho^*$, for positive constant $\rho^*  $. In the latter (former) case the corresponding eigenmodes possess {a} nontrivial (trivial) Chern number. The value $\rho^*$ represents a topological transition point.

Afterward, the edge problem is examined. Localized edge states are found on a semi-infinite domain. In the topologically nontrivial case, a family of gapless bands corresponding to edge states is found; the modes propagate unidirectionally. In the topologically trivial case the edge mode bands do not span the band gap and they can propagate bidirectionally.

\subsection{Bulk spectral bands}
\label{linear_bulk_bands}

Begin by considering the linearized discrete system {of} 
(\ref{HC_TB_eqn1})-(\ref{HC_TB_eqn2}) on an infinite domain. The spectral dispersion surfaces are found by taking Fourier solutions of the form 
\begin{align}
&a_{mn}(z) = A({\bf k},z) e^{ i {\bf k} \cdot (m {\bf w}_1 + n {\bf w}_2 )} \; ,\\ \nonumber
&b_{mn}(z) = B({\bf k},z) e^{ i {\bf k} \cdot (m {\bf w}_1 + n {\bf w}_2 )} \; ,
\end{align}
which yield 
\begin{equation}
\label{HC_spec_sys}
\frac{d {\bf c}}{dz} 
 =  i \mathcal{M}({\bf k},z) {\bf c} \; , ~~~~~ {\bf c} = \begin{pmatrix}
A \\ B
\end{pmatrix}({\bf k},z) \; ,
\end{equation}
such that
\begin{equation*}
\label{M_matrix_define}
\mathcal{M}({\bf k} ,z) = 
\begin{pmatrix}
0 & \tau({\bf k},z) \\
\tau({\bf k},z)^* & 0
\end{pmatrix} \; ,
\end{equation*}
for $ \tau({\bf k},z) = e^{i \theta_0(z) } + \rho \left( e^{i \theta_1(z) - i {\bf k} \cdot {\bf v}_1} + e^{i \theta_2(z) - i {\bf k} \cdot {\bf v}_2} \right)$. 
Notice that $\mathcal{M}({\bf k} ,z)^{\dag} = \mathcal{M}({\bf k} ,z)$, where $\dag$ denotes the conjugate transpose i.e. it is Hermitian. The corresponding reciprocal lattice vectors are given by
\begin{equation}
\label{BZ_vecs}
{\bf k}_1 = \frac{2 \pi}{3} \left( 1, \sqrt{3}\right) ,  ~~~~ {\bf k}_2 = \frac{2 \pi}{3} \left( 1, -\sqrt{3}\right) \; .
\end{equation}
We also note that the matrix $\mathcal{M}$ is $T$-periodic in $z$ and periodic in the spectral plane: $\mathcal{M}({\bf k} + p {\bf k}_1  ,z) = \mathcal{M}({\bf k} ,z ) = \mathcal{M}({\bf k} +q {\bf k}_2 ,z) $ where $p,q \in \mathbb{Z}.$

\begin{figure}
\includegraphics[scale=.36]{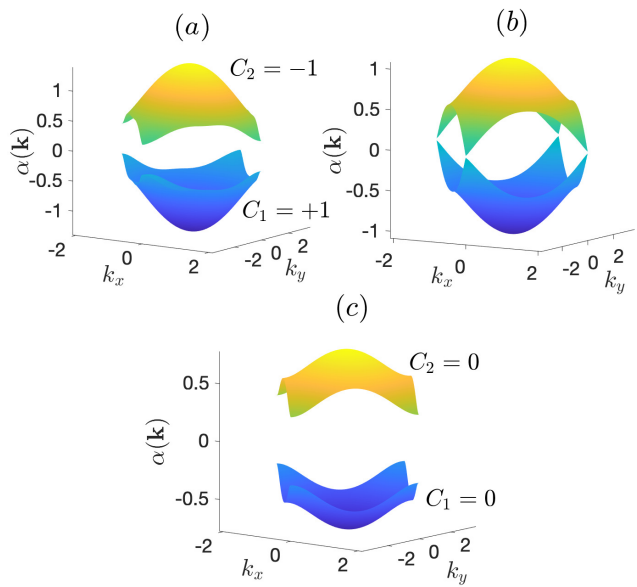}
 \caption{Bulk dispersion surfaces (\ref{floquet_define}) computed from (\ref{HC_spec_sys}) with $\kappa = 1.36$ and $\epsilon = 0.245$ at: (a) $\rho = 0.7$, (b) $\rho = 0.45$, and (c) $\rho = 0.2$. Included are the associated Chern numbers, where applicable, for the bands. 
 \label{spectral_surfaces}}
\end{figure}

Floquet modes and their corresponding Floquet exponents for system (\ref{HC_spec_sys}) are computed via Floquet theory \cite{Eastham73}. These solutions are assumed to satisfy the quasi-periodic boundary condition
\begin{equation}
{\bf c}({\bf k},z+T)  = e^{- i \alpha({\bf k}) T} {\bf c}({\bf k},z) ~ . 
\end{equation}
The monodromy matrix corresponding to (\ref{HC_spec_sys}) is found from the principal fundamental matrix solution at $z = T$. The eigenvalues of the monodromy matrix, $\gamma({\bf k})$, are known as the characteristic multipliers. The Floquet exponents are calculated directly by
\begin{equation}
\label{floquet_define}
\alpha({\bf k}) = \frac{i \log[\gamma({\bf k})] }{T}  \; ,
\end{equation}
from the fundamental branch.

For a typical set of values the 
spectral surfaces are plotted in Fig.~\ref{spectral_surfaces} over the unit cell {for} 
different values of $\rho$. In each case there are two spectral surfaces sorted in ascending order. For $\rho = 0.7$, there are two gaps present due to the periodicity in $\alpha$. {When $\rho > \rho^*$, the eigenmodes of the system possess nontrivial topological invariants.  As the value of $\rho $ decreases a topological transition takes place. The transition point  $\rho^* \approx 0.45$, where the central band gap closes, separates the nontrivial from trivial topological regions. When $\rho < \rho^*$, there is an open {central} gap, however the corresponding eigenmodes have trivial topological invariants. We point out that the model in \cite{Ablowitz14} was normalized differently and so it has a different transition point ($\rho^* \approx 0.5$).}




The topological invariant associated with this system is the Chern number. Nonreciprocal modes are {possible} 
with a nonzero Chern number. The 2D Chern number of eigenmode ${\bf c}_p$, corresponding to the $p^{\rm th}$ spectral band, is {given} by 
\begin{equation}
\label{chern}
C_p = \frac{1}{2 \pi i} \iint_{\rm UC} \left(  \frac{\partial {\bf c}_p^{\dag}}{\partial k_x}  \frac{\partial {\bf c}_p}{\partial k_y}- \frac{\partial {\bf c}_p^{\dag}}{\partial k_y}  \frac{\partial {\bf c}_p}{\partial k_x} \right) ~ d{\bf k} \; , ~~~ p = 1,2
\end{equation}
where UC denotes the unit cell (parallelogram) in ${\bf k}$-space defined in terms of the reciprocal lattice vectors in (\ref{BZ_vecs}). The Chern number is a $z$-invariant integer quantity of Eq.~(\ref{HC_spec_sys}) i.e. $C_p(z) = C_p \in \mathbb{Z}$ (see Appendix \ref{proof_chern_invariant}). The Chern numbers are numerically computed using the algorithm given in \cite{Fukui05}.


\subsection{Edge spectral bands}
\label{linear_edge_bands}

{We} now consider the  edge problem on a semi-infinite domain. Take the $y$ $(x)$-direction to be infinite (finite). We seek localized edge modes that decay exponentially fast in the direction perpendicular to the boundary. As such, we consider modes of the form
\begin{equation}
a_{mn}(z) = a_n({\bf k}, z) e^{ i {\bf k} \cdot m {\bf w}_1}  , ~~ b_{mn}(z) = b_n({\bf k}, z) e^{ i {\bf k} \cdot m {\bf w}_1}  ,
\end{equation}
which reduce (\ref{HC_TB_eqn1})-(\ref{HC_TB_eqn2}) to
\begin{align}
\label{HC_TB_1d_eqn1}
&  i \frac{d a_{n}}{dz} + [\mathcal{L}_+ b]_{n} + \sigma  |a_n|^2 a_n = 0 \; , ~~~ \\
\label{HC_TB_1d_eqn2}
 & i \frac{d b_{n}}{dz} + [\mathcal{L}_- a]_{n}+ \sigma  |b_n|^2 b_n = 0 \; ,
\end{align}
where ${\bf k} \cdot {\bf w}_1 = \frac{\sqrt{3}}{2} k_y$ and
\begin{align*}
 [\mathcal{L}_+ b]_{n}   =&  e^{i \theta_0(z)} b_{n}  \\ \nonumber
+ &  \rho \left[ e^{i \theta_1(z) - i {\bf k} \cdot {\bf w}_1 }  + e^{i \theta_2(z) + i {\bf k} \cdot {\bf w}_1 }  \right] b_{n-1} \; , \\
 [\mathcal{L}_- a]_{n}   =&  e^{-i \theta_0(z)} a_{n}  \\ \nonumber
+ &  \rho \left[ e^{-i \theta_1(z) + i {\bf k} \cdot {\bf w}_1 } + e^{-i \theta_2(z) - i {\bf k} \cdot {\bf w}_1 }  \right] a_{n+1} \; .
\end{align*}
The boundary conditions used are Dirichlet zero:
\begin{align}
\label{Dirichlet_BCs}
&b_n = 0 \; , ~~ {\rm for} ~~ n <0   , ~  n > N-1  ,  \\ \nonumber
&a_n = 0 \; , ~~ {\rm for} ~~ n <1   ,  ~ n > N  ,
\end{align}
(see Fig.~\ref{honey_lattice_fig} for reference). 

\begin{figure}
\includegraphics[scale=.4]{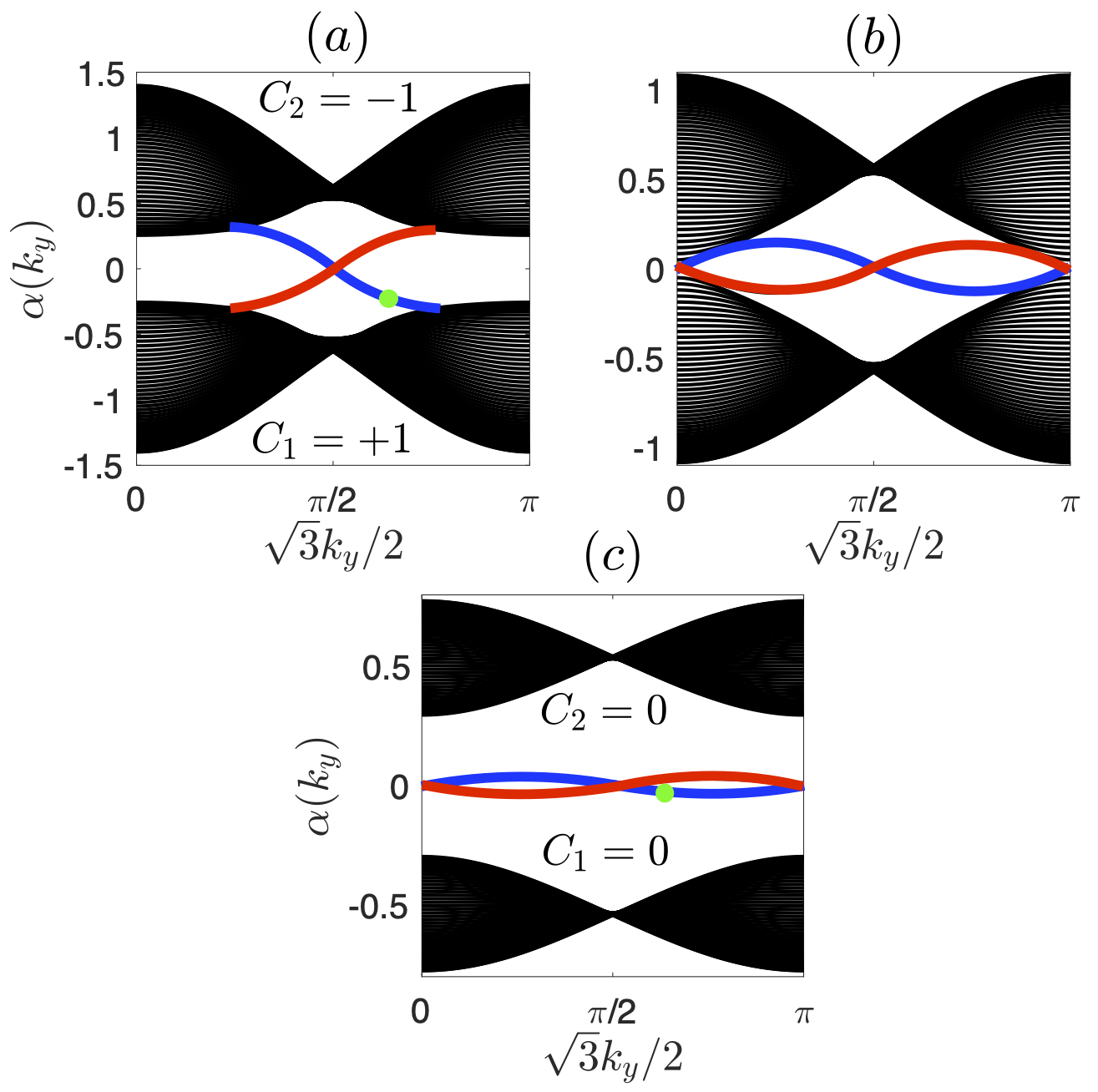}
 \caption{Linear edge Floquet bands computed from (\ref{HC_TB_1d_eqn1})-(\ref{HC_TB_1d_eqn2}) for: (a) $\rho = 0.7$, (b) $\rho = 0.45$, and (c) $\rho = 0.2$. The blue (red) bands correspond to edge modes localized along the left (right) domain wall. {Black regions correspond to bulk modes.} A green dot denotes the edge mode localized along the left edge with {mode number} 
 $k_y= 4/\sqrt{3}$.    \label{band_diagrams}}
\end{figure}

The linear ($\sigma = 0$) edge Floquet modes are computed in a similar manner to that of the bulk case above. Solutions are assumed to satisfy the quasi-periodic boundary condition
\begin{equation}
\begin{pmatrix}
a(k_y,z+T)  \\ b(k_y,z+T) 
\end{pmatrix}_{n}= e^{- i \alpha(k_y) T}
\begin{pmatrix}
a(k_y,z) \\ b(k_y,z)
\end{pmatrix}_n \; .
\end{equation}
 The monodromy matrix corresponding to (\ref{HC_TB_1d_eqn1})-(\ref{HC_TB_1d_eqn2}) is numerically computed  with the boundary conditions (\ref{Dirichlet_BCs}) imposed. The Floquet exponents  are computed from the characteristic multipliers, $\gamma(k_y)$, by 
\begin{equation*}
\alpha(k_y) = \frac{i \log[\gamma(k_y)] }{T}  \; ,
\end{equation*}
taking the fundamental branch.

Edge Floquet band diagrams, whose parameters {are the same as} those shown in Fig.~\ref{spectral_surfaces}, respectively, are displayed in Fig.~\ref{band_diagrams}. The solid black regions correspond to bulk eigenmodes. In each case the edge states (blue and red curves) are localized along the left and right domain walls. When $\rho  = 0.7 > \rho^*$ the edge state spans the gap and the corresponding group velocities are sign-definite. This is the topological case with nontrivial Chern numbers. Modes with negative (positive) slope correspond to negative (positive) group velocity {localized} along the left (right) boundary. The corresponding {(positive)} chiral mode is the combination of these two edge modes; it propagates counterclockwise, as viewed from the input, along the domain boundary.  


At $\rho^* \approx 0.45 $ a topological transition is indicated by the closing of the gap between the two bulk bands. Below this transition point (e.g. $\rho = 0.2 < \rho^*$) there are no gapless edge bands i.e. no {family of edge modes} 
 spans the entire gap.  {In this scenario, an envelope may travel in either direction (positive or negative $y$-direction) along the boundary wall. When a mode is propagating in this regime  and encounters a lattice defect along the boundary, significant backscattering can occur \cite{Ablowitz17}. }


{Topologically protected modes are identified through the bulk-edge correspondence. In the $\rho > \rho^*$ case, whose bulk bands in Fig.~\ref{band_diagrams} possess nontrivial Chern numbers, the  {edge states correspond} to a topologically protected mode. The upper bulk band has Chern number $C_2 = -1$, which {equals} 
the number of topological edge states in the gap above it (zero) minus the number of edge states in the gap below it (one). A similar algebra exists for the lower bulk band.

For the $\rho < \rho^*$ case, a pair of trivial Chern numbers is not sufficient to establish that these are topologically nontrivial. Indeed, when there {are} 
the same number of edge states above {the bulk band} as there are below, a (net) zero Chern number follows \cite{Ablowitz19}. This is not the case here, however, as the lack of edge modes above the upper bulk band indicates that the bidirectional edge state in the central gap must be topologically trivial.}



\section{Effective NLS Equation}
\label{asymptotics_sec}

Let us {now} consider waveguides that are rapidly rotating i.e. the inverse angular frequency is small: $0 < \epsilon \ll 1$. Furthermore assume a weakly nonlinear regime where $\sigma = \epsilon$.
A multiple scales analysis (see Appendix \ref{multi_scale}) reveals{, to leading order,} {edge} modes along the left boundary {in the neighborhood of $k_y=K$} of the form
\begin{equation}
\label{nonlinear_soln_asym}
a_{mn}(z) = 0 \; , ~~~~ b_{mn}(z) = C(y_m,z) r^n e^{i  {K} y_m } \; ,
\end{equation}
{where $y_m = \sqrt{3} m /2$ is considered as a continuous variable sampled at points on the discrete grid}
and $|r(K)| < 1$ indicates an exponentially decaying edge mode as $n \rightarrow \infty$. The slowly-varying envelope $C(y,z)$ satisfies the generalized NLS equation
\begin{align}
\nonumber
i \frac{\partial C}{\partial z} & +\alpha_*C+ i\alpha_*' \frac{\partial C}{\partial y} + \frac{\alpha_*''}{2} \frac{\partial^2 C}{\partial y^2} -  i \frac{\alpha_*'''}{6} \frac{\partial^3 C}{\partial y^3} + \dots \\ 
\label{gen_NLS_eqn}
& + \alpha_{\rm nl} |C|^2C + \dots = 0 \; ,
\end{align}
such that $\alpha_{\rm nl} =  \epsilon || b_n(K) ||_4^4/|| b_n(K) ||_2^2$ and $\alpha_*^{(p)} = \frac{d^p \alpha}{ dk_y^p} \big|_{k_y =K}$. A similar description exists on the right boundary where instead $a_{mn}$ is the nontrivial solution and $b_{mn}$ is approximately zero.


\begin{figure}
\centering
\includegraphics[scale=.37]{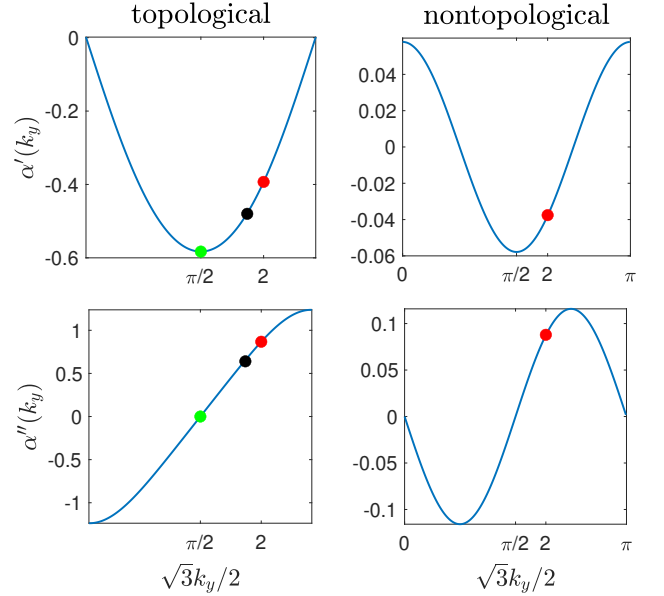}
 \caption{{Dispersion curves corresponding to edge modes in Fig.~\ref{band_diagrams} localized along the left boundary. (left column) $\rho = 0.7$, topological; (right column) $\rho = 0.2,$ nontopological. The green, black, and red dots denote the location of the $k_y = \pi/\sqrt{3}, ~ 2.18, $ and $4/\sqrt{3}$ modes, respectively.} \label{dispersion_bands_top}}
\end{figure}


For a  slowly varying envelope, the higher-order dispersion terms (beyond third-order) in (\ref{gen_NLS_eqn}) are {typically negligible}. 
For not too large power levels the higher-order nonlinearity terms (beyond cubic Kerr term) {are also} neglected. Certain linear terms {can be omitted by judiciously selecting the mode number, $K$, {and scaling}.} 
For instance,  near a maximum or minimum point of the edge dispersion bands in Fig.~\ref{dispersion_bands_top} {(red dot)}, $\alpha_*'' \not= 0$ and $\alpha_*''' \approx 0$. Hence the equation that governs the envelope dynamics is the traveling NLS equation 
\begin{equation}
\label{NLS}
i \frac{\partial C}{\partial z}  + \alpha_*C+ i \alpha_*' \frac{\partial C}{\partial y}  + \frac{\alpha_*''}{2} \frac{\partial^2 C}{\partial y^2} + \alpha_{\rm nl} |C|^2C = 0 \; ,
\end{equation}
 which, for $\alpha_*'' > 0$,  admits the traveling bright soliton solution {
\begin{equation}
\label{soliton_define}
C(y,z) = \nu \sqrt{\frac{\alpha_*''}{\alpha_{\rm nl}}} {\rm sech}\left[ \nu \left( y - \alpha_*' z \right) \right] e^{i \left( \frac{\alpha_*'' \nu^2}{2} +  \alpha_* \right) z}  ,
\end{equation}
}where $\nu \in \mathbb{R}$. {Note, {apart from a change of variables, this is equivalent to}
 the soliton solution in Eq.~(\ref{soliton_soln}).}
A comparison of this envelope with the full numerical solution of (\ref{HC_TB_eqn1})-(\ref{HC_TB_eqn2}) was explored in \cite{Ablowitz17}.

On the other hand, if we consider modes near an inflection point {[see Fig.~\ref{dispersion_bands_top} (green dot)]}, then $\alpha_*'' \approx 0$ while 
$\alpha_*'''  \not=  0$ and (\ref{gen_NLS_eqn}) reduces to the third-order NLS equation {
\begin{equation}
\label{higher_order_NLS}
i \frac{\partial C}{\partial z}  +\alpha_*C+ i \alpha_*' \frac{\partial C}{\partial y} - i\frac{\alpha_*'''}{6} \frac{\partial^3 C}{\partial y^3} + \alpha_{\rm nl} |C|^2C = 0 \; ,
\end{equation}
} which does not admit {any known} stable soliton modes.

The results of this section give us an analytic description of 
nonlinear modes {with slowly-varying spatial profiles}  in this rapidly-rotating asymptotic regime. The {solution form} 
in (\ref{nonlinear_soln_asym}) consists of a (possibly) topologically protected linear mode modulated by a slowly-varying envelope which satisfies the generalized NLS equation in (\ref{gen_NLS_eqn}). {Since these nonlinear modes} {modulate the underlying 
linear modes they inherit similar properties: unidirectionality and non-backscattering from defects.} 
{Since the modes in this regime are wide and nearly continuous,} { the effects of the PN barrier are {small}.} 

However{,} to obtain these results, we assumed {\it wide} spatial envelopes. 
The question we explore next is what happens when {narrower envelopes are considered as initial conditions (ICs). How does the PN energy barrier affect these ICs?} Moreover, {how does} the asymptotic results of this section shed light on the mode dynamics? The main finding, simply put, is that topologically protected modes never stop, meanwhile nontopological modes {slow down and eventually stop; they slow quickly when the modes} are narrow enough. Over  large enough distances the  discrete topological modes rearrange themselves{, approaching} a particular mode that is described by {the} asymptotic theory of this section.

\section{The Peierls-Nabarro Effect}
\label{PN_section}


In this section we study nonlinear {edge} modes {that initially} are not well approximated by the slowly-varying envelope theory of the previous section. Namely, we examine {a number of} highly-localized initial states and monitor their dynamics over long distances. The mode behavior is found to be directly related to the topology: topologically protected {edge} modes (with nonzero Chern number) do not slow down, whereas nontopological {edge} modes (with zero Chern number) can and do eventually stop translating. Furthermore, we point out this is a purely {\it nonlinear} effect; the PN barrier does not occur in the linearized problem, regardless of the associated topology.

\subsection{Soliton edge waves}

To investigate {the propagation of soliton edge waves} 
we initialize (\ref{HC_TB_eqn1})-(\ref{HC_TB_eqn2}) with envelopes of different sizes using the form in (\ref{nonlinear_soln_asym}) and (\ref{soliton_define}). The initial {edge wave} conditions taken are 
\begin{equation}
\label{define_IC}
\begin{pmatrix}
a_{mn}(0) \\ b_{mn}(0)
\end{pmatrix} = A_0 ~{\rm sech}\left[ \nu (y_m - y_*) \right]
\begin{pmatrix}
a_{n}(K) \\ b_{n}(K) 
\end{pmatrix} e^{iK y_m},
\end{equation}
where $A_0,\nu > 0 $, $y_*$ is the initial position,  and $y_m= \sqrt{3} m/2$ {is} the continuous variable, $y$, sampled at points on the discrete grid {for $m \in \mathbb{Z}$}. 
The eigenmodes $a_n,b_n$ are numerically calculated from the linearized version of (\ref{HC_TB_1d_eqn1})-(\ref{HC_TB_1d_eqn2}) with corresponding Floquet exponent values that lie in a band gap of Fig.~\ref{band_diagrams}.  In all simulations the linear mode is normalized  so that $b_0 = 1$ and we set $y_* =  (\sqrt{3}/2) ~224$, so that $m_* = 224$. We assume a weak nonlinearity and fix $\sigma = \epsilon = 0.245$ for all simulations below. 

When
\begin{equation}
\label{balanced} 
A_0 = \nu \sqrt{ \frac{\alpha_*''}{\alpha_{\rm nl}}} , 
\end{equation}
we {can} recover the soliton modes identified in the previous section when $0 < \nu \ll 1$ [see Eq.~(\ref{soliton_define})]; we refer to this as the {\it balanced} case. All simulations in this section correspond to balanced initial conditions.  In Sec.~\ref{off_balance_section} 
we examine {\it unbalanced} cases where $A_0 \not= \nu \sqrt{\alpha_*''/\alpha_{\rm nl}} $. The unbalanced case can be viewed as a perturbation of an ideal soliton balance. 

 The discrete system is integrated with a fourth-order Runge-Kutta method. We take a large computational domain in $m$ and implement periodic boundary conditions. The Dirichlet zero boundary conditions in (\ref{Dirichlet_BCs}) are imposed for $n$. An absorbing boundary layer described in Appendix \ref{ABL} is used to absorb and dissipate most radiation emitted from the edge state, as well as most of the reflections off the opposite edge wall. All measurements calculated below only involve the dominant mode, $b_{mn}$, since $a_{mn}$ is typically quite small.
 
  To begin, we track the location of the peak magnitude along the edge ($n = 0$) where $|b_{mn}(z)|$ is largest. This point is denoted by $m_{p}(z).$  The magnitude of $b_{mn}$ is found to always be largest near the edge. 

\subsection{{Topological mode propagation}}
\label{top_case_pn}

First, consider the topological case whose corresponding linear band diagram is shown in Fig.~\ref{band_diagrams} ($\rho = 0.7$). The corresponding dispersion curves are shown in Fig.~\ref{dispersion_bands_top}.  {We first} examine the edge mode localized along the left edge at $K  = 4/\sqrt{3}$ where $\alpha_*'' > 0$  and a wide envelope is dominated by second-order dispersion. {It is also clear} from the group velocity 
that all wave packets must move unidirectionally and in the negative direction since $\alpha'(k_y) < 0$.  {For} longitudinal length scales on the order $O(1/\epsilon)$ or less, a wide {($\nu \ll 1$)} but balanced envelope is {expected to be} effectively described by the focusing NLS equation in (\ref{NLS}).

\begin{figure}
\centering
\includegraphics[scale=.34]{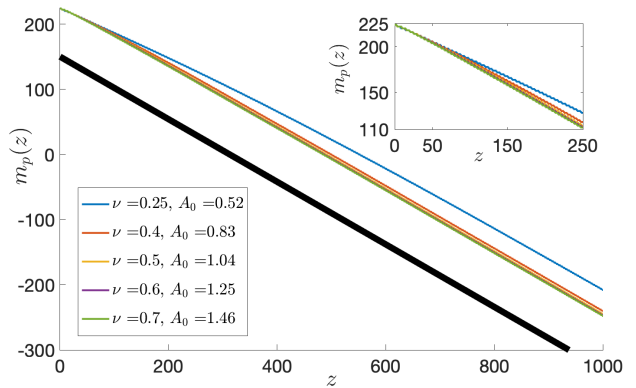}
 \caption{Location of soliton peak as a function of $z$ for the nonlinear, topological case. Balanced initial data (30) with different values of $\nu$ are studied.   {The thick black line is a trajectory with group velocity $\alpha' = -0.48$ corresponding to the mode {$k_y$}$ = 2.18$ in Fig.~\ref{dispersion_bands_top}. At fixed $z$,  curves appear top-to-bottom in increasing values of $\nu$. Inset: peak location for $0 \le z \le 250$. 
 }
 \label{topological_peakloc_fig}}
\end{figure}

Let us examine the location of the peak magnitude for various envelope widths. The results are shown in Fig.~\ref{topological_peakloc_fig}.
Regardless of the initial amplitude,  each realization is observed to translate {nearly uniformly} in the negative $y$-direction over very large distances {($z = 1000$).}

At intermediate distances ($0 \le z \le  250 \approx 1000 \epsilon $) the widest mode ($\nu = 0.25$) travels {approximately} with the {\it linear} group velocity ($\alpha'_* \approx -0.39$) of the mode corresponding to $ K = 4/\sqrt{3}$. However, at extremely long distances ($z > 500$) the nonlinear wave packet has a faster velocity of approximately $-0.48$. This observation is beyond the limit of the asymptotic theory in Sec.~\ref{asymptotics_sec}. On the other hand, {for} modes with narrow initial width (larger $\nu$), the modal velocity approaches the same group velocity of $-0.48$ almost immediately.

 We note a clear contrast between the mode dynamics observed here and those in 
 generic discrete nonlinear systems {(recall Fig.~\ref{DNLS_peak_pos})}. To move from one lattice site to another, a soliton must shed energy through radiation. {In DNLS (\ref{stdNLS})} 
  this occurs at the expense of the soliton velocity until the mode eventually ceases propagating. Here we {do not observe this; rather} the mode never slows down or stops. In other words, the inherent topology of the {system forces a mode to travel forward, in spite of the PN barrier.}

\subsection{Nontopological mode propagation}
\label{nontop_case_pn}

Now let us consider the {edge wave}  behavior when the corresponding linear band diagrams are nontopological [see Fig.~\ref{band_diagrams} ($\rho = 0.2$)]. Recall that this topological transition, from nontrivial to trivial,  is achieved by deforming the individual waveguides in a preferred direction, as adjusted through the  parameter $\rho$.  Examining the band structure in Fig.~\ref{band_diagrams}, as $\rho$ decreases the bands tend to flatten out, resulting in overall smaller values of $\alpha$ and its derivatives. The group velocity in the nontopological case is about a tenth of the size of the topological one, and so the dynamics take longer to develop{, again $K=4 /\sqrt{3}$.} 

\begin{figure}
\centering
\includegraphics[scale=.34]{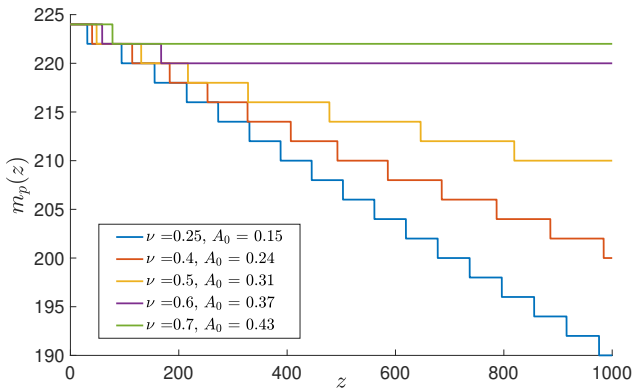}
 \caption{Location of soliton peak as a function of $z$ for the nonlinear, nontopological case. Balanced initial data (30) with different values of $\nu$ are studied. At fixed $z$,  curves appear top-to-bottom in decreasing values of $\nu$.
 \label{peakloc_nontop_fig}}
\end{figure}

Numerical results are shown in Fig.~\ref{peakloc_nontop_fig}. There is a clear difference in these nontopological mode dynamics from {the topologically protected case.} 
In the nontopological case the solitons {\it do} suffer from PN barrier {slow-down} effects. As the amplitude increases (width decreases) the soliton slows 
from its initial velocity and ultimately stops translating completely. The energy imbalance of the PN barrier causes the mode to transition from a propagating state to a stationary one. This behavior resembles the classic {mode dynamics observed} 
in the DNLS equation {(see Fig.~\ref{DNLS_peak_pos})}. {We also  remark that at lower amplitude values, as with the DNLS equation, it takes a very long time to see the slow down effect.} The effect of a nontrivial Chern number in highly discrete 
systems is apparent: {\it{localized modes with nontrivial topology, unlike their nontopological counterparts, never cease propagating.}} 

\begin{figure}
\centering
\includegraphics[scale=.34]{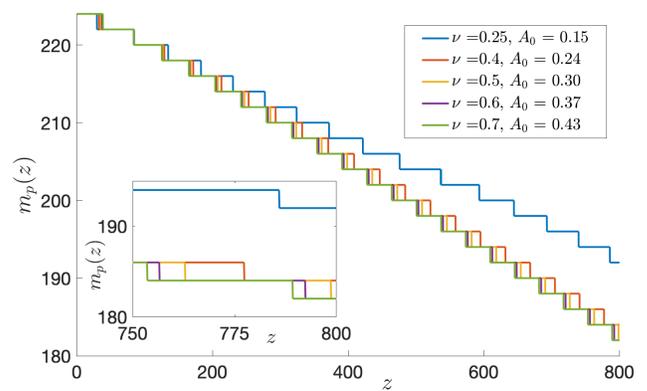}
 \caption{Location of soliton peak as a function of $z$ for the linear, nontopological case. The initial conditions are the same as 
 in Fig.~\ref{peakloc_nontop_fig}.  At fixed $z$,  curves appear top-to-bottom in increasing values of $\nu$. Inset: peak location for $750 \le z \le 800$.
 \label{peakloc_nontop_lin_fig}}
\end{figure}

\subsection{Linear edge waves}
\label{nontop_case_pn}
Finally, we  reiterate that the PN barrier is fundamentally a {\it nonlinear} 
phenomenon.} To highlight this we repeat the simulations using the same initial conditions of Fig.~\ref{peakloc_nontop_fig}, but now we remove the nonlinear terms from the governing equations (\ref{HC_TB_eqn1})-(\ref{HC_TB_eqn2}) i.e. we set $\sigma = 0.$ 
These {edge} modes are topologically trivial states (zero Chern number). The results are presented in Fig.~\ref{peakloc_nontop_lin_fig}. Unlike the nonlinear dynamics, here all modes are found to propagate freely. Hence the results found above are physically important for high-power (nonlinear) beam propagation in  Kerr media. We note that the widest mode ($\nu = 0.25$) travels at the linear group velocity corresponding to $K = 4/\sqrt{3}$, while the other modes travel faster.

{But we also note that linear modes are not uniform traveling waves;  over long distances, they suffer from dispersion.} {A standard stationary-phase calculation \cite{Ablowitz_Dispersive_Book} shows that envelopes governed by the linear ($\alpha_{\rm nl} = 0$) generalized NLS Eqn.~(\ref{gen_NLS_eqn}) decay like {$z^{-1/2}$} for (\ref{NLS}) or {$z^{-1/3}$} for (\ref{higher_order_NLS}).}

\section{Long distance states}
\label{long_time_sec}

The large $z$ dynamics of initially well-localized, nearly discrete, topological {edge} modes are studied in this section. {Even though the topology of the system forces the propagation of modes, there are still PN barrier effects. These effects {are manifest} in the form radiation shedding from which a localized mode {is observed to} redistribute its spatial profile from discrete (narrow) to effectively continuous (wide) states where the {effects of the} PN barrier {are smaller}.
Upon transitioning to this effectively continuous state, we observe that the profile appears to be  governed by the generalized NLS equation {(\ref{gen_NLS_eqn})}{; this is} discussed in Sec.~\ref{asymptotics_sec}.}

To study the modal evolution, two quantities are monitored: the peak magnitude along the left edge (at $n = 0$),
\begin{equation}
\label{magnitude}
A(z) = \max_{m} |b_{m0}(z)| , 
\end{equation}
with $A(0) = A_0$, and the participation number, 
\begin{equation}
\label{part_num}
P(z) = \frac{ \left( || b_{m0}(z) ||^2_2 \right)^2 }{ || b_{m0}(z) ||^4_4} ,
\end{equation}
where $|| b_{m0} ||_p = \left( \sum_{m} |b_{m0}|^p \right)^{1/p}$. The location of the soliton peak, $m_p(z)$, was already examined in Sec.~\ref{PN_section}. The participation number is a quantity  which is inversely proportional to the width of a localized mode i.e. small $P(z)$ corresponds to well-localized modes and vice versa. {The participation number is an indirect way of measuring the  width.}

To develop some intuition for the participation number, consider approximating sums by integrals: $\sum_m \Delta y \approx \int  dy$. This approximation is reasonable as long as our function is slowly-varying relative to the grid spacing. 
Then the participation number  is approximated by
\begin{align*}
P(z) &= \frac{1}{  \Delta y} \frac{ \left( \sum_m |b_{m0}(z)|^2 \Delta y \right)^2 }{ \sum_m |b_{m0}(z) |^4 \Delta y} \\ 
&\approx \frac{1}{\Delta y} \frac{ \left( \int_{-\infty}^{\infty} |C(y,z)|^2 dy \right)^2}{ \int_{-\infty}^{\infty} |C(y,z)|^4 dy}  ,
\end{align*}
where  $C(y,z)$ is a continuous function that approximates the profile of $b_{mn}(z)$ parallel to the edge. For the secant hyperbolic function in (\ref{soliton_define}), $\int |C(y,z)|^2 dy = \frac{2\nu (\alpha_*'')^2}{ \alpha_{\rm nl}^2}  $ and $\int |C(y,z)|^4 dy =\frac{4 \nu^3 (\alpha_*'')^4}{3 \alpha_{\rm nl}^4}$, and so the (constant) soliton participation number for $\nu \ll 1$ is
\begin{equation}
\label{sol_part_num}
P_{\rm soliton}(z) =  \frac{\sqrt{3}}{ \nu} ,
\end{equation}
{ where we recall {$y_m=\sqrt{3}m/2$ so that $\Delta y=y_{m+2} - y_m = \sqrt{3}$}.} 

There is a large parameter space in which to search. To begin, we focus on balanced initial modes at a particular modal value  
and then gradually increase the localization i.e. increase $\nu$. {Afterward, results  for {some} 
off-balanced initial conditions and their resulting dynamics are discussed.} 

\subsection{{Topological mode propagation -- balanced case}}
\label{wide_mode_dynamics}

We {first} 
study the modal evolution of wide {and} balanced soliton {edge wave} states. 
The position of the peak magnitude for these modes was {examined} 
in Sec.~\ref{PN_section}A. The peak magnitude and participation number dynamics measured for this case 
{are} shown in Fig.~\ref{top_balanced_mode_evolve}. As the following modes reorient their structure, they radiate waves. In all cases, the mode profile retains a localized edge form despite a potentially significant radiation loss.

\begin{figure}
\centering
\includegraphics[scale=.3]{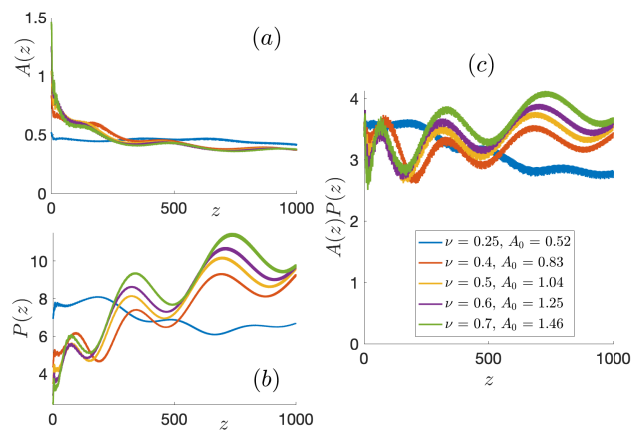}
 \caption{(a) Peak magnitude, (b) participation number, and (c) product of peak magnitude and participation number evolutions for the balanced case. Initial data (\ref{define_IC}) with $A_0 = \nu \sqrt{\alpha_*''/\alpha_{\rm nl}}$ {and $K = 4/\sqrt{3}$} is used.  At  $z = 1000$,  curves in panel (c) appear top-to-bottom in decreasing values of $\nu$. \label{top_balanced_mode_evolve}}
\end{figure}

Consider a relatively wide profile ($\nu = 0.25$) {with an initial {wave} number $K = 4/ \sqrt{3}$ corresponding to positive dispersion ($\alpha''(K) > 0$) in Fig.~\ref{dispersion_bands_top} (left column)}. The parameter evolutions in Fig.~\ref{top_balanced_mode_evolve}, for values  of $z$ well beyond order $O(1/\epsilon)$, show that the modal dynamics are  well described by the theory in Sec.~\ref{asymptotics_sec}: the magnitude and width do not substantially change from their initial low amplitude and wide profile input. Due to the wide, slowly-varying nature of the spatial envelope, PN barrier effects are minimal.

{It is natural to} {expect these modes to} effectively be solitons with the same form as the secant hyperbolic function in (\ref{soliton_define}), where the amplitude and width are related by {(\ref{balanced}).} 
If this is the case, then the product between the amplitude and the participation number [recall (\ref{sol_part_num})] should be steady-state and equal $A(z) P(z) =  \sqrt{\frac{ 3 \alpha_*''}{\alpha_{\rm nl}}} $ which, for these parameters, is approximately $3.60$. Examining the numerical results in Fig.~\ref{top_balanced_mode_evolve}(c),  we obtain a value at  $z = 250 $ of $A(z) P(z) \approx 3.53$, a difference of $1.94 \%$. At a long-distance of $z = 1000 $ we find $A(z) P(z) \approx 2.80$, a difference of $22.22 \%.$ Clearly at very large distances, well beyond the range of the asymptotic theory in Sec.~\ref{asymptotics_sec}, higher-order dispersive terms eventually become non-neglible and affect the mode.

To further establish the solitonic nature of these modes, we apply a secant hyperbolic fit to the edge profile, the result of which is shown in Fig.~\ref{topological_balance_fit} (left column). The close agreement between the functional form in (\ref{soliton_define}) and numerical output is noteworthy.  The wide envelope ($\nu = 0.25$) case is observed to lose some amplitude without much change in width.  To obtain the fit we numerically measure the peak magnitude and participation number. To get $\nu$, we apply formula (\ref{sol_part_num}) {with the numerically computed participation number}. Up to $z = 250$ the mode should be governed by a self-focusing NLS type equation [like (\ref{NLS})] on a background. At large $z$ values, higher-order dispersive effects become significant; hence at $z  = 1000$ we note the presence of a small dispersive tail on the trailing (right) edge of the solitary hump. 

\begin{figure}
\centering
\includegraphics[scale=.35]{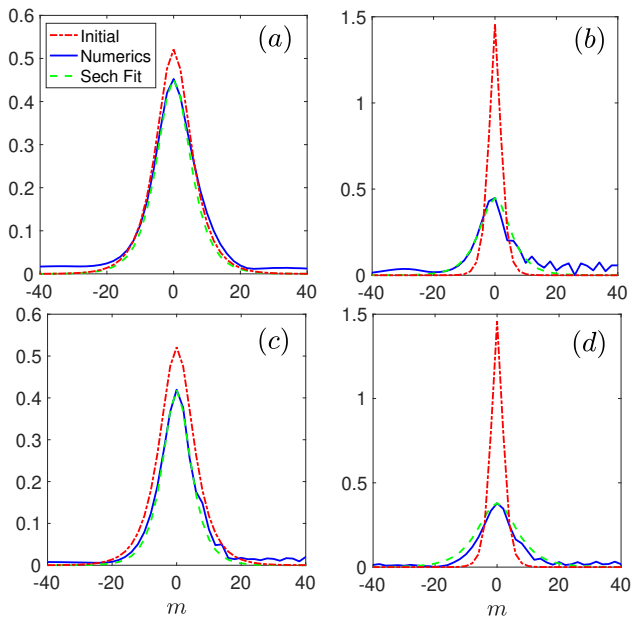}
 \caption{Initial condition ($z = 0$), numerical solution at given $z$, and secant hyperbolic (Sech) fit of  edge profile magnitude, $|b_{m0}(z)|$, for the topologically protected, balanced case. The initial width $\nu$ and $z$ value of the profiles are: (a) $(\nu,z) = (0.25, 250)$, (b) $(\nu,z) = (0.7, 250)$, (c) $(\nu,z) = (0.25, 1000)$, and (d) $(\nu,z) = (0.7, 1000)$. Note: each mode shown here has been re-centered to $m = 0$ to  appreciate their structure.
 \label{topological_balance_fit}}
\end{figure}

Now let us examine {balanced} {{edge} modes that are initially} narrow. 
Recall that  the velocities of these wave packets were already examined in Fig.~\ref{topological_peakloc_fig}; there each of the modes in this narrow width regime {was} 
approaching a particular 
{envelope} state {as} identified by its group velocity. At first glance, no part of this discussion is within the framework of the asymptotic theory described in Sec.~\ref{asymptotics_sec}. However,  we observe a transition from {one that is initially narrow} 
to a wide one that actually appears to be described by our asymptotic results.

The evolution of the peak magnitude [Fig.~\ref{top_balanced_mode_evolve}(a)] and participation number [Fig.~\ref{top_balanced_mode_evolve}(b)]  for the $\nu = 0.7$ mode reveals a precipitous decrease in the amplitude accompanied by a decrease in  localization ($P(z)$ increase). The product between these two quantities is {oscillating, but eventually is approaching} 
a {nearly} steady-state value of $A(z) P(z) \approx 3.69$, indicating an approximate solitonic state.

The profile of the edge mode is shown in Fig.~\ref{topological_balance_fit} (right column). From this it is clear that the, initially, large amplitude and narrow width mode transitions to one that is smaller and wider, effectively continuous. When we apply a secant hyperbolic fit, similar to the manner above, we observe a reasonable agreement. A feature of these modes is a dispersive tail on the (right) trailing tail of the envelope. This indicates that even at early distances ($z\approx 250$) there are higher-order dispersion effects present that have a noticeable impact on the mode profile, unlike the initially wide mode.

{Next we discuss how to} reconcile these findings with the asymptotic results of Sec.~\ref{asymptotics_sec}. Once the mode is wide, {our numerics indicates that it is well described by } 
{the} generalized NLS equation in (\ref{gen_NLS_eqn}). Hence, to good approximation, the mode is traveling with the linear group velocity of approximately $-0.48$ (see Fig.~\ref{topological_peakloc_fig}). Consulting the linear dispersion curves in Fig.~\ref{dispersion_bands_top}, it appears that the {modulated envelope} 
state corresponds to $k_y \approx 2.18$ {[see Fig.~\ref{dispersion_bands_top} (black dot)]}. This mode number is closer to the inflection point, $k_y = \pi/\sqrt{3} \approx 1.81$ {[see Fig.~\ref{dispersion_bands_top} (green dot)]}, where the effective envelope equation is governed by the third-order NLS equation (\ref{higher_order_NLS}). Moreover, the second and third order dispersion coefficients are nearly equal at this point, with $\alpha'' / 2 \approx 0.33 \approx \alpha''' / 6$. Hence this ultimate mode represents an intermediate state {of (\ref{gen_NLS_eqn}), where the second and third order coefficients are balanced.} 
{The next coefficient in the expansion, $\alpha^{(4)}/24$, is considerably smaller and can be effectively neglected.}

\subsection{Off-balance Modal Evolutions}
\label{off_balance_section}

In {the previous section} 
the Peierls-Nabarro effect was explored for {\it balanced} nonlinear modes. This investigation followed naturally from the asymptotic analysis of Sec.~\ref{asymptotics_sec} which described a family of balanced soliton states. 
{Here} we  explore 
some modes that are {\it unbalanced} and compare them to their balanced counterparts. Overall, we do not observe any fundamental differences between these states and the balanced modes considered above. Indeed, the {ability of topologically protected modes to resist a slow down due to the PN energy barrier is a robust feature.} 

\begin{figure}
\centering
\includegraphics[scale=.5]{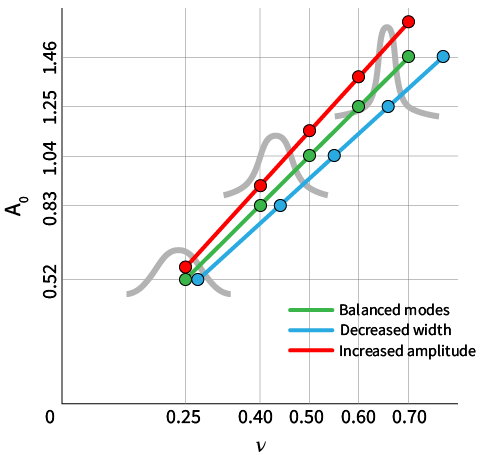}
 \caption{Initial amplitude and width used in initial condition (\ref{define_IC}).  The center green line corresponds to  the balanced initial state. The upper red (lower blue) line corresponds to increased amplitude (decreased width) relative to the balanced cases. The off-balanced {case is computed by increasing $A_0$ ($\nu$) {from} the green line while leaving $\nu$ ($A_0$) fixed.} Points on the center green, upper red, and lower blue lines correspond to the results shown in Figs.~\ref{top_balanced_mode_evolve}, \ref{top_amp_increase_mode_evolve}, and \ref{top_nu_increase_mode_evolve}, respectively. 
 \label{IC_map}}
\end{figure}

To organize all the modes that we will be considering, a phase diagram 
is given in Fig.~\ref{IC_map} {{which describes} the initial amplitude-width {combinations} 
we will consider.} 
In all cases below we take {wave number} $K = 4 / \sqrt{3}$. As a result, the balanced line has a slope of  $\sqrt{\alpha_*'' / \alpha_{\rm nl}} \approx 2.08$, where $\alpha_*''$ and $\alpha_{\rm nl}$ were defined in (\ref{gen_NLS_eqn}). To obtain an off-balanced case we introduce  a ten-percent perturbation away from the (green) balanced line: {\it decreased width} (increased $\nu$) corresponds to a positive shift $\nu \rightarrow 1.1 \nu $, for fixed $A_0$; meanwhile {\it increased amplitude} (increased $A_0$) is the positive shift $A_0 \rightarrow 1.1 A_0 $, where $\nu$ is fixed.

\begin{figure}
\centering
\includegraphics[scale=.32]{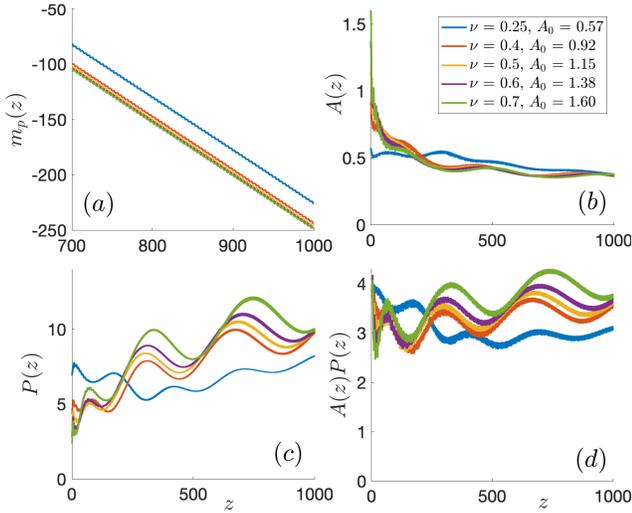}
 \caption{(a) Peak position, (b) peak magnitude, (c) participation number, and (d) product of peak magnitude and participation number evolutions for the unbalanced, increased amplitude case (red line in Fig.~\ref{IC_map}). Initial data (\ref{define_IC}) with $ A_0 = 1.1 \nu \sqrt{\alpha_*''/\alpha_{\rm nl}} $  is taken. At  $z = 1000$,  curves in panels (a) and (d) appear top-to-bottom in increasing and decreasing values of $\nu$, respectively.
  \label{top_amp_increase_mode_evolve}}
\end{figure}

{First,} consider the case where the amplitude is increased relative to the balanced state, but the width is left the same (red line in Fig.~\ref{IC_map}). The dynamics of edge states corresponding to different values of $\nu$ are summarized in Fig.~\ref{top_amp_increase_mode_evolve}. Each mode {considered} is found to translate with {a velocity of approximately $-0.48$. Recall this is the same velocity observed in the balanced narrow modes of Fig.~\ref{topological_peakloc_fig}.} 
Apparently, the larger amplitude has perturbed all states into the {same state ($k_y \approx 2.18$) as was identified in {the previous section.} 

In the course of evolution, each of {these} localized modes loses significant amplitude and widens. Moreover all modes considered here appear to be approaching the same state at large $z$. The magnitude and participation evolutions in Figs.~\ref{top_amp_increase_mode_evolve}(b-d) resemble those for the balanced case in Fig.~\ref{top_balanced_mode_evolve} with larger $\nu$. Hence, at large $z$, this mode is approximately a solitonic state with some {minor} higher-order dispersion effects present. Furthermore, we observe that the amplitude-participation product in Fig.~\ref{top_amp_increase_mode_evolve}(d) is nearly the same value as the narrow modes in Fig.~\ref{top_balanced_mode_evolve}(c). This indicates a rather stable system: small amplitude perturbations of the initial conditions only leads to small perturbations of the resulting dynamics.  

\begin{figure}
\centering
\includegraphics[scale=.34]{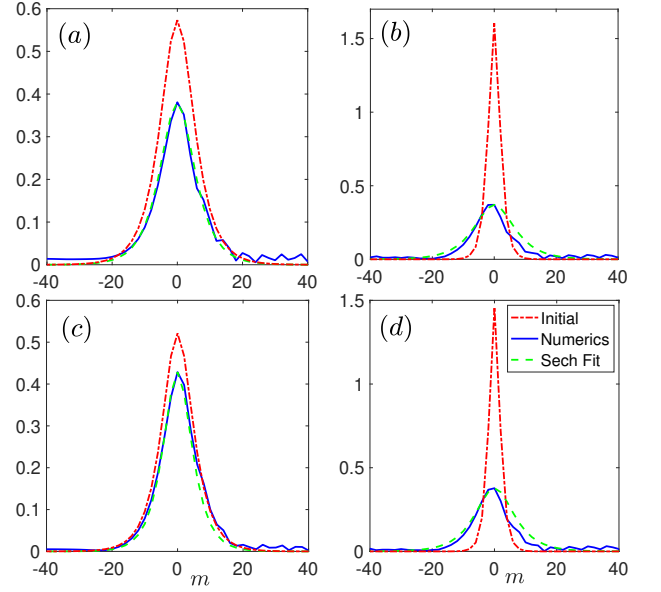}
 \caption{Initial condition ($z = 0$), numerical final condition ($z = 1000$), and secant hyperbolic (Sech) fit of  edge profile magnitudes, $|b_{m0}(z)|$, for the topologically protected, but unbalanced cases. The initial width $\nu$ and amplitude $A_0$ of the profiles are: (a) $(\nu,A_0) = (0.25, 0.57)$, (b) $(\nu,A_0) = (0.7, 1.60)$, (c) $(\nu,A_0) = (0.275, 0.52)$, and (d) $(\nu,A_0) = (0.77, 1.46)$.  {Top row:} increased amplitude case; {Bottom row:} decreased width case. {Left column:} wider mode; {Right column:} narrower mode. Note: each mode shown here has been re-centered to $m = 0$ to help {describe} their structure.
 \label{offbalance_amp_fit}}
\end{figure}

Edge mode profile {comparisons} are shown in Fig.~\ref{offbalance_amp_fit} (top row) for two cases. A secant hyperbolic function fit is applied, in the same manner as in Fig.~\ref{topological_balance_fit}. The final states shown correspond to very large propagation distances and each is observed to be reasonably well approximated by a secant hyperbolic function, despite a dispersive (trailing) tail that is clearly forming. Again, this suggests that these modes are governed by an effective NLS equation in which 
{the third-order dispersive effects} 
 play a non-negligible role {over large distances}.

\begin{figure}
\centering
\includegraphics[scale=.33]{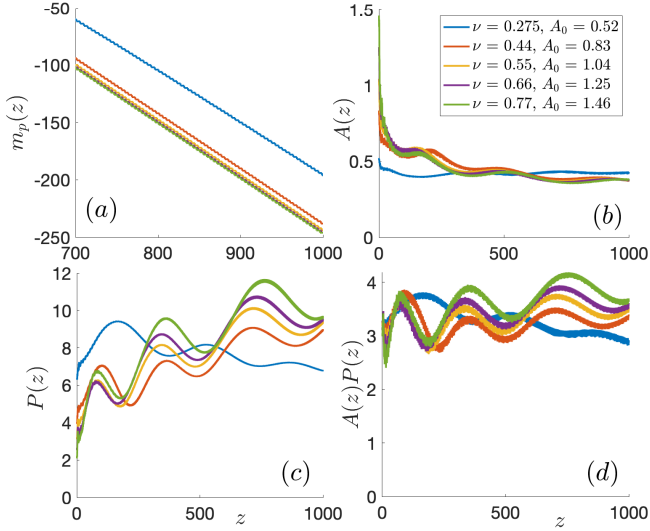}
 \caption{(a) Peak position, (b) peak magnitude, (b) participation number, and (c) product of peak magnitude and participation number evolutions for the unbalanced, decreased width case (blue line in Fig.~\ref{IC_map}). Initial data (\ref{define_IC}) with $ \nu = 1.1 A_0 \sqrt{\alpha_{\rm nl}/\alpha_*''} $ is taken. At  $z = 1000$,  curves in panels (a) and (d) appear top-to-bottom in increasing and decreasing values of $\nu$, respectively.
 \label{top_nu_increase_mode_evolve}}
\end{figure}

Now let us consider when the initial form (\ref{define_IC}) is narrower ($\nu$ larger) relative to the balanced case, but amplitude remains unchanged (see blue line in Fig.~\ref{IC_map}). The {resulting} mode dynamics are shown in Fig.~\ref{top_nu_increase_mode_evolve}. The envelope velocities observed  in Fig.~\ref{top_nu_increase_mode_evolve}(a)  resemble those in the balanced case (see Fig.~\ref{topological_peakloc_fig}): the narrowest modes travel faster than then wide mode. The narrow modes ($\nu \ge 0.44$) travel with the velocity identified {several times mentioned above} 
($-0.48$) and the widest mode ($\nu = 0.275$) is slightly slower.

Like above, Figs.~\ref{top_nu_increase_mode_evolve}(b-c) indicate that a mode sheds some radiation and find{s a nonlinear envelope mode governed by the generalized NLS equation (\ref{gen_NLS_eqn}).} 
Again, the amplitude and participation product of Fig.~\ref{top_nu_increase_mode_evolve}(d) is nearly constant with small oscillations and indicates there is nearly a steady-state solitonic mode with some higher-order dispersion. 

The profiles for the widest and narrowest initial states considered are displayed in Fig.~\ref{offbalance_amp_fit} (bottom row). The wider state has a small dispersive tail and is approximated reasonably well by a secant hyperbolic function. The narrow mode has a significant oscillating tail and is not as well approximated by a hyperbolic secant profile. Nonetheless, our asymptotic theory here continues to {be useful} 
for these unbalanced states.

\subsection{{Topological} {radiating traveling wave states}}
\label{hamil_steady_strong_perturb_sec}

{In these next} simulations we examine extremely localized {edge} modes in the topological regime.} {We  fix the initial amplitude in (\ref{define_IC}) and substantially} reduce the width; i.e. fix $A_0$, but increase $\nu \gg 1$. {The purpose of these runs is to explore what happens to very narrow modes localized to nearly one lattice site.} The position of the solitary wave peak is shown in Fig.~\ref{peak_location_strong_perturb}. {As {$\nu$ increases} 
we see 
that the modes have larger speed {and tend to a particular speed. Said differently, as the initial localization increases, the modes appear to approaching a special state.} 

The {empirically} measured group velocity of the modes is found to be {nearly} that of the $k_y = \pi/\sqrt{3}$ mode in Fig.~\ref{dispersion_bands_top} {(green dot)}. {This  {mode} 
corresponds to a localized envelope 
which slowly modulates the linear edge mode at the inflection point of the dispersion band. As a result, the effective envelope equation is the third-order NLS equation in (\ref{higher_order_NLS}). {What is interesting about this finding is that these modes have the largest group velocity of any admissible edge state; this behavior is different from what was observed in the nontopological case where energy is transferred to a stationary mode (smallest group velocity magnitude).}

{We also note that, while not shown here, these modes lose a substantial amount of amplitude through radiation and develop long dispersive tails.  {We 
ascribe this} to the PN barrier.  
{We also note that the third-order NLS (\ref{higher_order_NLS}) is not known to support stable soliton modes.}
{Since they are associated with substantial radiation, the}  modes found here are not 
{as well-suited as balanced  states} 
for transporting coherent structures over long distances.}

\begin{figure}
\centering
\includegraphics[scale=.38]{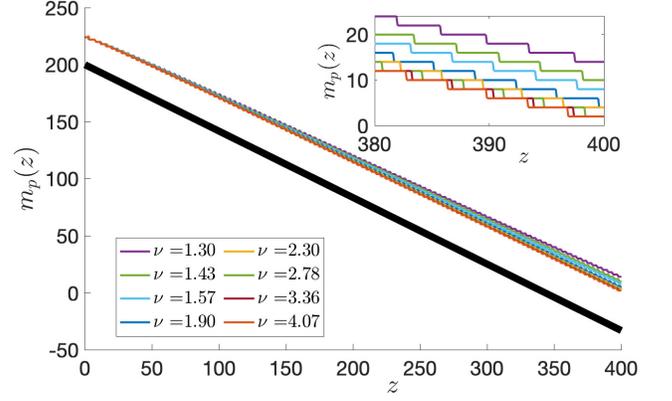}
 \caption{{Location of soliton peak as a function of $z$ for the nonlinear, topological case.  Initial data (\ref{define_IC}) with $A_0 = 1.04$ and different values of $\nu$ {are used.} All other parameters are the same as  Fig.~\ref{topological_peakloc_fig}. {The thick black line is a trajectory with group velocity $\alpha' = -0.583$ corresponding to the mode $k_x = \pi/\sqrt{3}$ ({green dot)} in Fig.~\ref{dispersion_bands_top}.} At fixed $z$,  curves appear top-to-bottom in increasing values of $\nu$. Inset: peak location for $380 \le z \le 400$. 
 \label{peak_location_strong_perturb}}}
\end{figure}

{In our final calculation, we {consider} the Hamiltonian of the system.} 
The Hamiltonian was introduced in \cite{Ablowitz15}, it reads}
\begin{align}
\nonumber
H = - \sum_{m,n} \Big( &a_{mn}^*  \left[ \mathcal{L}_+  b \right]_{mn} + b_{mn}^*  \left[ \mathcal{L}_-  a \right]_{mn} \\ \label{Hamil_define} &+ \frac{\sigma}{2} |a_{mn}|^4 + \frac{\sigma}{2} |b_{mn}|^4 \Big) .
\end{align}
{The governing system of equations (\ref{HC_TB_eqn1})-(\ref{HC_TB_eqn2}) are obtained from the Hamiltonian system
\begin{equation}
i \frac{d a_{mn}}{dz} = \frac{\delta H}{\delta a_{mn}^*} , ~~~~~ i\frac{d b_{mn}}{dz} = \frac{\delta H}{\delta b_{mn}^*} ,
\end{equation}
where $i a_{mn}, a_{mn}^*$ and $i b_{mn}, b_{mn}^*$ form two sets of conjugate variables, and $\delta/\delta x$ denotes {the} 
standard variational derivative with respect to $x$.}
Many authors have used energy arguments e.g. Hamiltonians, to explain the PN barrier effect \cite{Kivshar93,Jenkinson16} {(see also Sec.~\ref{background_subsection})}. {In particular, when a mode suffers from the PN barrier, it ultimately settles at the state which minimizes the Hamiltonian.}
In general, {the Hamiltonian here} {is} 
$z$-dependent, so we  focus instead on the average Hamiltonian {quantity}
\begin{equation}
\label{ave_hamil_define}
H_{\rm ave} = \frac{1}{T} \int_0^T H(z) dz .
\end{equation}

The average Hamiltonian for a wide and balanced nonlinear envelope {of the form in (\ref{define_IC}) 
is shown in Fig.~\ref{top_disperse_vs_hamil}. This is done to give an indication of a preferred state.
The  inflection point $(k_y = \pi/\sqrt{3})$ is not a minimum of the averaged Hamiltonian; actually no critical minimum point (i.e. $dH_{ave}/ dk_y = 0$ and $d^2H_{ave}/ dk_y^2 > 0$) is found. Instead the inflection point corresponds to a critical  maximum point where  $d H_{\rm ave} / d k_y = 0$ and $d^2H_{ave}/ dk_y^2 < 0$.  The states corresponding to slower group velocity are found to have negative  Hamiltonian values that are rapidly decreasing.} 
{A full description of these states in terms of the Hamiltonian is beyond the scope of this paper.}

\begin{figure}
\centering
\includegraphics[scale=.38]{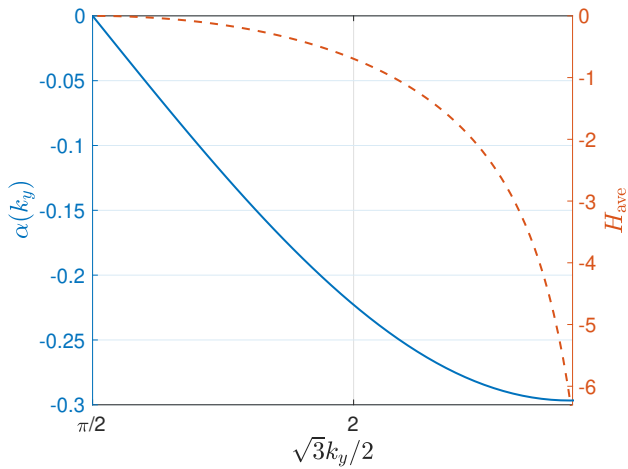}
 \caption{{The average Hamiltonian value (dashed curve, {right axis}) of a {typical wide} {($\nu = 0.25$)} balanced, and topologically protected envelope initial profile (\ref{define_IC}). Also included is a portion of the {topologically protected} edge dispersion  band (solid curve, {left axis}), originally shown in Fig.~\ref{dispersion_bands_top}.
 \label{top_disperse_vs_hamil}}}
\end{figure}

\section{Conclusions}
\label{conclude_sec}

 The Peierls-Nabarro barrier is a classic nonlinear phenomenon {often} found in 
 discrete nonlinear {equations; it}  becomes significant when the width of a mode is on the order of the lattice spacing. In nontopological systems this barrier {can cause discrete} {traveling solitary waves} to slow down and eventually stop i.e. discrete traveling waves {would not be} supported. {In this case it is found that} {highly localized solitary waves quickly slow down and eventually stop. For less localized traveling waves the time to stop can be exponentially long.}
 
{The main result in this paper is that we find nonlinear topologically protected modes (i.e. those with nonzero Chern number) do not suffer from the PN barrier in the usual way. Nonlinear topological modes are found to always continue propagating unidirectionally, meanwhile they 
widen and adjust their amplitudes to approach slowly-varying wave envelopes that are governed by a generalized NLS equation in the large distance limit.}

{The results of this paper have important practical implications. Current Floquet topological insulator systems are fabricated using femtosecond waveguide etching techniques that lie in the tight-binding limit where discrete models, such as those   studied above, are  good approximations. Many researchers desire highly localized modes centered almost totally at a single lattice site. For the parameter regimes studied above, our findings  indicate {that due to the PN barrier}
{an edge} mode will in general decrease amplitude and increase width through the shedding of radiation so that it can travel through the lattice.}}

{The field of nonlinear topological insulators is  still relatively new. Nonlinearity can lead to the existence of {soliton edge modes which satisfy the classical NLS equation; they} inherit features of the underlying linear topology of the system. This is an intriguing combination that admits robust topologically protected edge solitons. In addition to the topological protection, balanced and nearly balanced nonlinear modes remain close to soliton states over long distances. 
Strongly off-balanced modes are found to approach radiating traveling wave states that satisfy a third-order NLS equation. We find that the extremum of the average Hamiltonian 
is useful in describing this state. Finally, linear modes travel and decay due to dispersive/radiative effects over long distances.}

\section*{Acknowledgements}
\label{Acknowledge}
{This work was partially supported by AFOSR under Grant
No. FA9550-19-1-0084 and NSF under {Grants}
No. DMS-1712793 {and DMS-2005343.}}

\appendix

\section{$z$-invariant Chern Numbers}
\label{proof_chern_invariant}

The edge modes considered in this paper are categorized as topological when the adjacent bulk bands have a nonzero Chern number (\ref{chern}). 
{Here we show} that the Chern number{ $C$,} is a $z$-independent quantity{, i.e.}, $\frac{d C }{dz} = 0.$ 

To begin we make the change of variable in ${\bf k}-$space: $k_x = (k_u + k_v)/3, k_y = (k_v - k_u)/\sqrt{3}$ which maps the parallelogram defined by reciprocal lattice vectors ${\bf k}_1, {\bf k}_2$ in (\ref{BZ_vecs}) to a square of length $2 \pi.$ The form of the Chern number (\ref{chern}) is invariant under this transformation, so 
\begin{equation}
\label{chern2}
C = \frac{1}{2 \pi i} \iint_0^{2\pi} \left(  \frac{\partial {\bf c}^{\dag}}{\partial k_u}  \frac{\partial {\bf c}}{\partial k_v}- \frac{\partial {\bf c}^{\dag}}{\partial k_v}  \frac{\partial {\bf c}}{\partial k_u} \right)  dk_u dk_v \; ,
\end{equation}
 {corresponding to an eigenmode} ${\bf c}(k_u,k_v).$ Under the transformation, $ \tau({\bf k},z) = e^{i \theta_0(z) } + \rho \left( e^{i \theta_1(z) } e^{- i k_v } + e^{i \theta_2(z) } e^{- i k_u} \right)$ and $\mathcal{M}(k_u + 2\pi ,k_v  ,z) = \mathcal{M}(k_u,k_v,z)  = \mathcal{M}(k_u  ,k_v + 2\pi ,z) $. Differentiating (\ref{chern2}) with respect to $z$ {and utilizing Eq.~(\ref{HC_spec_sys}) yields 
\begin{align}
\label{dCdz_eqn}
\frac{d C}{dz} = \frac{1}{2 \pi} \iint_0^{2\pi} &\bigg( {\bf c}^\dag \frac{\partial \mathcal{M}}{\partial k_u}  \frac{\partial {\bf c}}{k_v}    - \frac{\partial {\bf c}^\dag}{\partial k_v} \frac{\partial \mathcal{M}}{\partial k_u} {\bf c} \\ 
\nonumber &-  {\bf c}^\dag \frac{\partial \mathcal{M}}{\partial k_v}  \frac{\partial {\bf c}}{k_u}    + \frac{\partial {\bf c}^\dag}{\partial k_u} \frac{\partial \mathcal{M}}{\partial k_v} {\bf c} \\ 
\nonumber &  + 2 \frac{\partial {\bf c}^\dag}{\partial k_u}  \mathcal{M} \frac{\partial {\bf c} }{\partial k_v}  - 2 \frac{\partial {\bf c}^\dag}{\partial k_v}  \mathcal{M} \frac{\partial {\bf c} }{\partial k_u} \bigg) dk_u dk_v .
\end{align}
}{Using} ${\bf c}(k_u + 2\pi, k_v, z) = {\bf c}(k_u, k_v, z) = {\bf c}(k_u, k_v + 2\pi, z)$ {and applying} integration-by-parts on (\ref{dCdz_eqn}), along with $\frac{\partial^2 \mathcal{M}}{\partial k_u \partial k_v} = 0,$ gives the desired result. Hence, 
{the} Chern numbers calculated in (\ref{chern}) are independent of $z$ {and furthermore can be computed from the eigenmode ${\bf c}(k_u,k_v, z)$ at any $z$}.

\section{Perturbation Analysis}
\label{multi_scale}

This appendix serves to highlight the main ideas in deriving envelope equation (\ref{gen_NLS_eqn}). A more detailed derivation can be found in \cite{Ablowitz14, Ablowitz17}. Starting from the edge problem  in (\ref{HC_TB_1d_eqn1})-(\ref{HC_TB_1d_eqn2}), a rapid variable is introduced, $\zeta = z/\epsilon$, in the presence of weak nonlinearity, $\sigma = \epsilon $, where $0 < \epsilon \ll1.$ The edge functions are expanded in powers of $\epsilon$
\begin{equation}
a_n = \sum_{j=0} \epsilon^j a_n^{(j)}(k_y,z,\zeta)  , ~~~ b_n = \sum_{j=0} \epsilon^j b_n^{(j)}(k_y,z,\zeta) ,
\end{equation}
and substituted into (\ref{HC_TB_1d_eqn1})-(\ref{HC_TB_1d_eqn2}). We look for leading-order solutions along the left zig-zag edge of the form $a_n^{(0)} = 0$ and $b_n^{(0)} = \widehat{C}(k_y,z) b_n^S(k_y)$, where  $\widehat{C}$ is a continuous spectral envelope and $b_n^S$ is an exponentially decaying mode of the form $b_n^S \sim  r^n$, satisfying $\overline{[\mathcal{L}_+ b^S]}_n = 0$. The overbar notation indicates the $z$-averaged quantity {$\overline{(~~)} = \frac{1}{2 \pi} \int_0^{2 \pi} (~~) d\zeta $}. The intervals of $k_y$ in Fig.~\ref{band_diagrams} where localized edge modes exist correspond to $|r(k_y)| < 1$. The slowly-varying envelope is found to satisfy the spectral equation
\begin{equation}
i \frac{d \widehat{C}}{dz} - \alpha(k_y) \widehat{C} + \alpha_{\rm nl}(k_y) | \widehat{C}|^2 \widehat{C} = 0 \; ,
\end{equation}
where $\alpha(k_y)$ is the Floquet exponent and $\alpha_{\rm nl}(k_y) = \epsilon || b_n(k_y) ||^4_4 / || b_n(k_y) ||_2^2$. The discrete norms are defined by 
\begin{equation*}
||f_n||_s =  \left( \sum_{n = 0}^{\infty} |f_n|^s \right)^{1/s} ~ , ~~~~ s = 2,4 ~ .
\end{equation*}
The linear and nonlinear coefficients are Taylor series expanded about $k_y = K$ 
\begin{equation*}
\alpha(k_y) = \alpha_* + \alpha'_* (k_y - K) + \frac{\alpha''_*}{2} (k_y - K)^2 + \cdots  ,
\end{equation*}
where $\alpha^{(p)}_* = \frac{d^p \alpha}{d k_y^p} \big|_{k_y = K} $ and $\alpha_{\rm nl}(k_y) = \alpha_{\rm nl}(K) + \cdots .$   A highly-localized spectral envelope 
is taken about a mode whose corresponding value of $\alpha(K)$ resides in the band gap. Applying the inverse Fourier transform in the linear terms
\begin{equation}
\label{space_env_fourier}
C(y,z) \approx \int_{\mathbb{R}} \widehat{C}\left( \frac{k_y -K}{\nu},z \right) e^{i k_y y} d k_y \; , ~~~ \nu \ll 1 \; ,
\end{equation}
{implies we can} replace $(k_y - K)$ with $-i \partial_y$. In order to excite a single mode edge state, the spectral envelope $\widehat{C}$ is taken to be highly localized about the point $k_y = K$. As a consequence, the spatial envelope in (\ref{space_env_fourier}) has a slowly-decaying form. 

\section{Absorbing Boundary Layer}
\label{ABL}

Highly nonlinear input modes tend to 
shed significant amounts of radiation in the first moments of their evolution. This radiation eventually reflects back off the edge of the computational domain and {then 
interacts with the main wave.}
To remove {the bulk of} this radiation, we implement an absorbing boundary layer to damp the functions $a_{mn}, b_{mn}$ for $1 \ll N_A \le n.$ In the 
numerical simulations presented in this paper we add the terms $  i \delta_n a_{mn}$ and $ i \delta_n b_{mn}$ to the left-hand side of Eqs.~(\ref{HC_TB_eqn1}) and (\ref{HC_TB_eqn2}), respectively, where
\begin{equation}
\delta_n =
\begin{cases}
0 & n < N_A \\
e^{- \eta} \left(  e^{\xi (n-N_A)} - 1  \right)  & n \ge N_A
\end{cases} \; ,
\end{equation}
is a  non-negative function with $\xi > 0$. Adding these terms has the effect of damping everything in the far-field region $n \ge N_A$. Since the function is not a perfect absorber, some reflection is expected. To mitigate this we take $\xi $ to be small{ so that the dissipation ramps up slowly.} In our simulations we chose {$\xi = 3 \times 10^{-3} , \eta = 2 \times 10^{-2},$} and $N_A = 100.$ 



\end{document}